\documentclass{aa}

\usepackage{natbib}
\usepackage{txfonts}
\usepackage[dvips]{graphicx}
\bibpunct{(}{)}{;}{a}{}{,}
\usepackage{booktabs,aalongtable}

\begin{document}

\title{VLBI observations of seven BL Lac objects from RGB sample}

\author{Zhongzu Wu \thanks{e-mail: zzwu@shao.ac.cn} \inst{1,2,3}, D. R. Jiang
\inst{1,2}, Minfeng Gu \inst{1,2}, Yi Liu \inst{1,2}
       }

\institute{ Shanghai Astronomical Observatory, Chinese Academy of
Sciences, Shanghai 200030, China
        \and
        Joint Institute for Galaxy and Cosmology (JOINGC) of SHAO
and USTC.
        \and
        Graduate School of the Chinese Academy of Sciences,
Beijing 100039, China
        }

\date{Received / Accepted}


\abstract{

We present EVN observations of seven BL Lac objects selected from
the RGB sample. To investigate the intrinsic radiation property of
BL Lac objects, we estimated the Doppler factor with the VLA or
MERLIN core and the total 408 MHz luminosity for a sample of 170 BL
Lac objects. The intrinsic (comoving) synchrotron peak frequency was
then calculated by using the estimated Doppler factor. Assuming a
Lorentz factor of 5, the viewing angle of jets was constrained. The
high-resolution VLBI images of seven sources all show a core-jet
structure. We estimated the proper motions of three sources with the
VLBI archive data, and find that the apparent speed increases with
the distance of components to the core for all of them. In our BL
Lacs sample, the Doppler factor of LBLs is systematically larger
than that of IBLs and HBLs. We find a significant anti-correlation
between the total 408 MHz luminosity and the intrinsic synchrotron
peak frequency. However, the scatter is much larger than for the
blazar sequence. Moreover, we find a significant positive
correlation between the viewing angle and the intrinsic synchrotron
peak frequency. The BL Lac objects show a continuous distribution on
the viewing angle. While LBLs have a smaller viewing angle than that
of IBLs and HBLs, IBLs are comparable to HBLs. We conclude that the
intrinsic synchrotron peak frequency is not only related to the
intrinsic radio power (though with a large scatter), but also to the
viewing angle for the present sample.

 }

\authorrunning{Z. Z. Wu et al.}
\titlerunning{VLBI observation of seven BL Lac Objects}

\maketitle

\section{Introduction}
BL Lac objects are a subclass of radio loud active galactic nuclei
(AGNs) with no emission lines or less of them, but they have a
strong continuum ranging from radio to $\gamma$-ray bands. Their
spectral energy distribution (SED) consists of a synchrotron
component at lower frequencies and an inverse Compton component at
higher frequencies. Historically, BL Lac objects have been
separately discovered through radio or X-ray surveys and have been
divided into two classes, namely, radio-selected BL Lacs (RBLs) and
X-ray selected BL Lacs (XBLs). However, this classification is based
on the observing band rather than on the intrinsic physical
properties. As a matter of fact, some BL Lac objects can be
classified as both RBL and XBL, such as Mrk 501. \cite{padovani95}
have suggested dividing RBL-like and XBL-like objects into
low-energy peaked BL Lacs (LBLs) and high-energy peaked BL Lacs
(HBLs) according to the location of the peak of the synchrotron
emission $\nu_{\rm peak}$ \citep{urry95}. Generally, RBLs tend to be
LBLs, and XBLs tend to be HBLs. The synchrotron peak frequency of
RBLs is usually in the radio/IR band, while UV/X-ray band for XBLs
\citep{Giommi95}. In recent years, samples including intermediate BL
Lac objects (IBLs) have been found in surveys that combine X-ray and
radio observations \citep{laure99,perl98,landt01,cacci99}.
Their discovery has shown that BL Lac objects most likely form one
class with a continuous distribution of synchrotron emission peak
energies, while RBLs and XBLs represent the opposite ends of the
continuum \citep{nieppola2006}.

The so-called `blazar sequence' was proposed to link the shape of
the SED and the synchrotron peak frequency to the source luminosity
for blazars, which consists of more luminous flat-spectrum radio
quasars (FSRQs) and BL Lac objects (Fossati et al. 1998; Ghisellini
et al. 1998). The most powerful sources have relatively small
synchrotron peak frequencies, and the least powerful ones have the
highest $\rm \nu_{peak}$ values. This anti-correlation can be
theoretically explained by the cooling processes. In more powerful
sources, the energy density is higher and the emitting particles
have a higher probability of losing energy so are subjected to more
cooling, resulting in a lower value for $\rm \nu_{peak}$. However,
this sequence is based on the absence of high-luminosity HBLs and
low-luminosity LBLs, therefore at least part of this systematic
trend can result from selection effects (Ant\'{o}n \& Browne 2005).
Indeed, the evidence of low-power LBLs has been recently discovered
(Padonavi et al. 2003; Caccianiga \& March\~{a} 2004; Ant\'{o}n \&
Browne 2005), and the possible discovery of high luminosity HBLs is
also reported in the Sedentary survey \citep{giommi05}. In addition,
the high-power-high-$\rm \nu_{peak}$ FSRQs were found in the Deep
X-ray Radio Blazar Survey (DXRBS), which is both X-ray and
radio-selected, though they do not reach the extreme $\rm
\nu_{peak}$ values of HBLs. From all these discoveries, it seems
that the blazar sequence in its simplest form cannot be valid
(Padovani 2006).

Instead of focusing on one or two limited surveys,
\cite{nieppola2006} have examined the properties of BL Lacs through
constructing the SEDs for a large, heterogeneous sample of BL Lacs
taken from the Veron-Cetty \& Veron BL Lac catalogue and visible
from the Metsh\"{a}hovi radio observatory. To our knowledge, this is
the largest BL Lacs sample to revisit the blazar sequence up to now,
though it is heterogeneous. The authors have found an
anti-correlation between the radio power and $\rm \nu_{peak}$, with
a huge scatter, reaching 5 orders of magnitude in power, and many
outliers were also found, especially in the low-power-low-$\rm
\nu_{peak}$ region. However, it is well known that the radio power
of BL Lacs is affected by the beaming effect of radio jets, since BL
Lacs are generally believed to have a relativistic jet aligned close
to the line of sight (Urry \& Padovani 1995). As the blazar sequence
was originally proposed as the anti-correlation between the
intrinsic luminosity and peak frequency, it nevertheless might be
important to revisit the correlation using the intrinsic radio
power. Moreover, the peak frequency $\rm \nu_{peak}\propto \it B \rm
\delta \gamma_{peak}^{2}$, where $B$ is the magnetic field, $\delta$
the Doppler factor, and $\rm \gamma_{peak}$ a characteristic
electron energy that is determined by a competition between
accelerating and cooling processes. While blazars as a whole
population suffer the Doppler boosting, the difference in Doppler
factor can exist between various subsets, which implies the
importance of investigating the intrinsic peak frequency after
excluding the Doppler factor. As \cite{koll} find, the average
angles of radio jets to the line of sight is approximately
$20^{\circ}$ for XBLs, but $10^{\circ}$ for RBLs. When the BL Lac
objects show a continuous distribution of synchrotron emission-peak
energies and most recent works focus on the correlation between the
power and peak frequency, it is not clear whether they also show a
continuous distribution for the angle to the line of sight, and HBLs
and LBLs occupy the opposite ends of this distribution.

Apart from investigating the SED of BL Lac objects, the
high-resolution radio observation is an important way to explore the
radio structure of BL Lacs, from which we can obtain the physical
information, such as the viewing angle, Lorentz factor and magnetic
field. From the VLBI observations, the different subsets of BL Lacs
can be compared based on the compact radio structure. In particular,
it enables us to explore the jet orientation for BL Lac objects,
because the properties of the parsec scale structure are strongly
dependent on jet orientation. \cite{koll96a} show that the
jets in XBLs fade more quickly than in RBLs. 
The VLBI observations show that most of LBLs display the rapid
superluminal apparent motions \citep{jorstad01}, while the TEV
blazar sources (most are HBLs) display subluminal or mildly
relativistic \citep{gir04b,gir06,pin04}. \cite{rec03} have observed
15 HBLs and 3 BL Lacs from the RGB sample and find that the HBLs,
like most LBLs, show parsec-scale core-jet morphologies with complex
kilo-parsec scale morphologies. Moreover, the jets in HBLs are more
well-aligned, suggesting that the jets of HBLs are either
intrinsically straighter or are seen further off-axis than LBLs.
\cite{gir04a} selected 30 low-redshift BL Lac objects and confirmed
that parsec and kilo-parsec scale jets are oriented at the same P.A.
in a large fraction of HBLs. The HBLs show less distortion and
therefore are expected to be oriented at larger angles than the LBL
sources. 

\cite{laure99} present a sample of 127 BL Lac objects from the ROSAT
ALL-Sky Survey-Green Bank catalogue (RGB), which exhibits properties
intermediate between high- and low-energy-peaked BL Lac objects.
In this paper, we present EVN observations of seven BL Lac objects
selected from the RGB sample. Our goal is to investigate
their compact radio structure. 
In combination with the previous observations and the archive data,
we aim to explore the possible structure variation and measure the
proper motion for our sources. Moreover, to explore BL Lac objects
as a whole population, we investigated the distribution of the
intrinsic radio power, the Doppler factor, and the viewing angle
along the intrinsic synchrotron peak frequency for a sample of BL
Lac objects selected from \cite{nieppola2006}.

The layout of this paper is as follows. In Sect. 2, the observations
and data reduction for our seven sources are given. The compact
radio structure are investigated in Sect. 3, where the proper motion
are estimated for three sources as well. In Sect. 4, the intrinsic
radio low-frequency luminosity, the Doppler factor, and the viewing
angle are estimated for a sample of BL Lacs selected from
\cite{nieppola2006}, and their distribution with the intrinsic
synchrotron peak frequency is also explored. The discussions are
shown in Sect. 5, while the conclusions are drawn in Sect. 6.
Throughout this work, we adopt the following values for the
cosmological parameters: $H_{0}=70 \rm {~km ~s^ {-1}~Mpc^{-1}}$,
$\rm \Omega_{M}=0.3$, and $\rm \Omega_{\Lambda} = 0.7$, except an
otherwise stated. The spectral index $\alpha$ is defined as $f_{\nu}
\propto \nu^{-\alpha}$.

\section{VLBI OBSERVATIONS AND DATA REDUCTION}\label{vlbiobs}

All seven BL Lac objects were observed with EVN at 5 GHz on 22
November 1999. The raw data were obtained with the MKIII VLBI
recording system, with an effective bandwidth of 28 MHz, and
correlated in Bonn at the Max-Planck-Institute. To study the
structure and proper motion of these sources, we collected EVN and
VLBA archive data. All seven sources are listed in Table
\ref{table1}: Column (1) Source name, Column (2) redshift, Columns
(3) the observational date and (4) the half-power beamwidth (HPBW)
of the weighted beam, Column (5) the noise of the image, Column (6)
the peak flux of the image, Column (7) VLBI array. All the data were
reduced using the NRAO Astronomical Image Processing System (AIPS)
package. The system temperatures and gains were obtained from the
VLBA website. After the initial reduction (including editing,
amplitude calibration, instrumental phase corrections, and
antenna-based fringe-fitting), we imported the calibrated data into
DIFMAP package (Shepherd, Pearson \& Taylor 1994, 1995) to make
images and fit the sources with a number of discrete circular
Gaussian components. The fitting was done directly on the final,
self-calibrated visibility data. The modelfit parameters are listed
in Table \ref{tab:model}.

\section{Results }\label{result}
The VLBI images show core-jet structure for all sources we selected,
and some of them are in good agreement with previous observations.
The MERLIN archive images and available VLA images also show that
these sources posse very straight jets. Model fitting of the radio
structures in each image was performed, using the MODELFIT of
Caltech DIFMAP package (uvweight=0, -1). The modelfit parameters are
presented in Table \ref{tab:model}. To determine the uncertainty of
the component position, we used the Difwrap package (Lovell 2000).
The errors were estimated by perturbing the fit parameters until the
resulting residual maps were unacceptable and taking the most
extreme accepted points. A visual inspection was done to determine
the goodness of the fit, and we used the best-fit total $\chi^2$ as
the upper limit for every fit of the components. 

In our seven BL Lac objects, previous VLBI observations at 6 cm are
available for five of them. Combining our observations with the
archive data, we got at least four-epoch observations for three
sources: 1727+502, 1133+704, and 1741+196. The weighted linear fits
to the component distances from the core as a function of the
observing time were used to estimate the proper motion of the
components.
The results are given in Table \ref{table2}, in which Column (1) is
the source name, (2) the components
label, (3) the proper motion, and (4) the apparent speed. 
From Table \ref{table2}, we find that all three sources have
superluminal motion, and the apparent speed increases with distance
from the core. For 1011+496 and 1424+240, only three-epoch data are
available, therefore we do not measure their proper motion and only
present their images in this work. For 4C+37.46 and 1542+614, there
are no previous VLBI images and archive data available, so we
present their first image.

\subsection{Comments on individual sources} \label{comment}

\begin{figure}[t]

 \resizebox{3in}{!}{\includegraphics{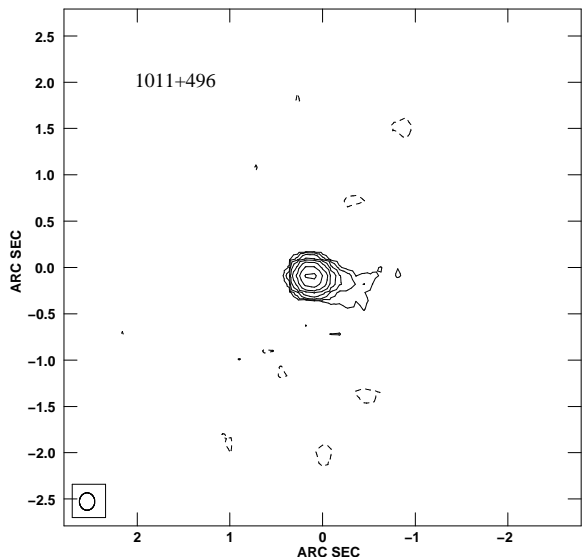}}
\caption{The MERLIN archive image of 1011+496 at 1.6 GHz on May 6,
1999. The contour levels are -1, 1, 2, 4, 8, 16, 32, 64... times
0.001742 Jy $beam^{-1}$(3$\sigma$). }\label{1011merlin}
\end{figure}

\begin{figure*}[]
 \centering
 {\includegraphics[width=12cm]{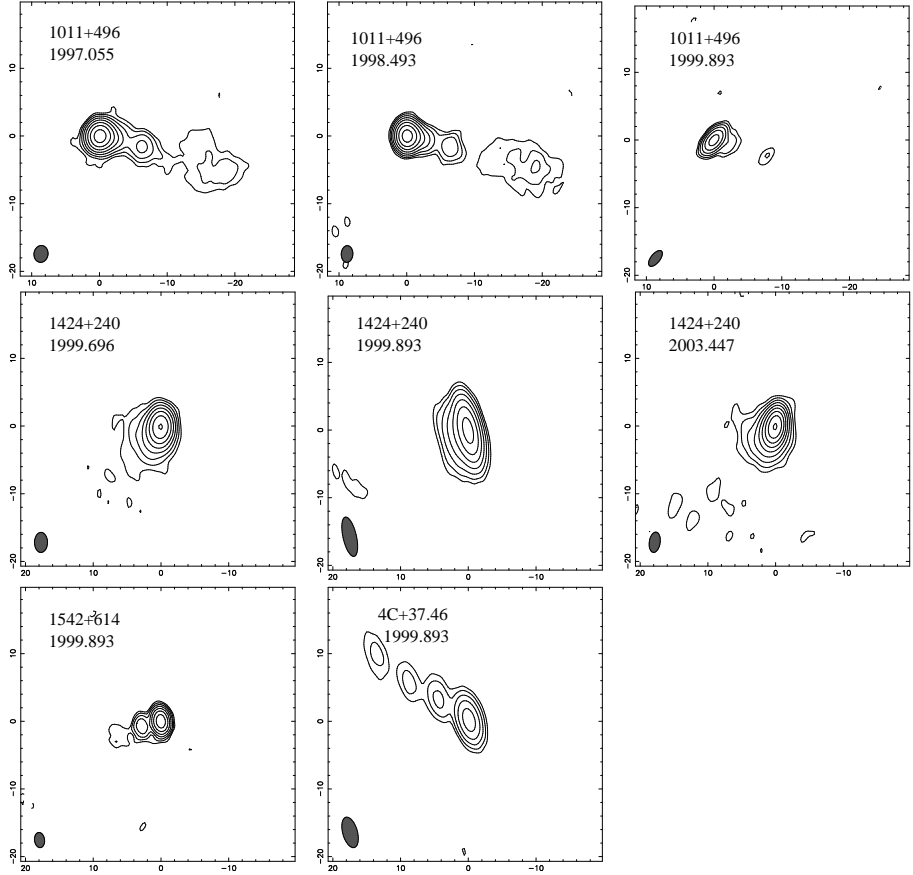}}
\caption{Multi-epoch EVN and VLBA images at 5 GHz. The minimum
contour level is 3 times the rms noise given in Table. \ref{table1}.
The contour levels are multiples(-1, 1,2,4,8,16,32,64...) of the
minimum contour level and the axes are labeled in milliarcseconds.
The top three panels are for 1011+496 at 1997.055, 1998.493, and
1999.893 (from left to right); the middle for 1424+240 at 1999.696,
1999.893, and 2003.447 (from left to right); while the bottom for
1542+614 and 4C+37.46 at 1999.893.}\label{otherall}
\end{figure*}

{\it 1011+496 --} Snapshot VLA observations have found hints of
extended structure around a bright core \citep{mach83}. This was
confirmed by \cite{koll}, who found diffuse emission extending
around $10^{\circ}$ to the north from their VLA B-configuration
data. \cite{AWB98} have modeled this source with three components,
which clearly showed that the jet of this source extends to about a
few hundreds mas of the core with position angle ~$-99^{\circ}$. The
MERLIN archive image (see Fig. \ref{1011merlin}) shows that the jet
extends in nearly the same direction with our EVN and VLBA images
(see Fig.\ref{otherall}), which show a jet in the opposite direction
on the parsec scale. As the jet direction is similar in MERLIN, EVN,
and VLBA images, this source is well-aligned.

\begin{figure*}
\resizebox{\hsize}{!}{\includegraphics{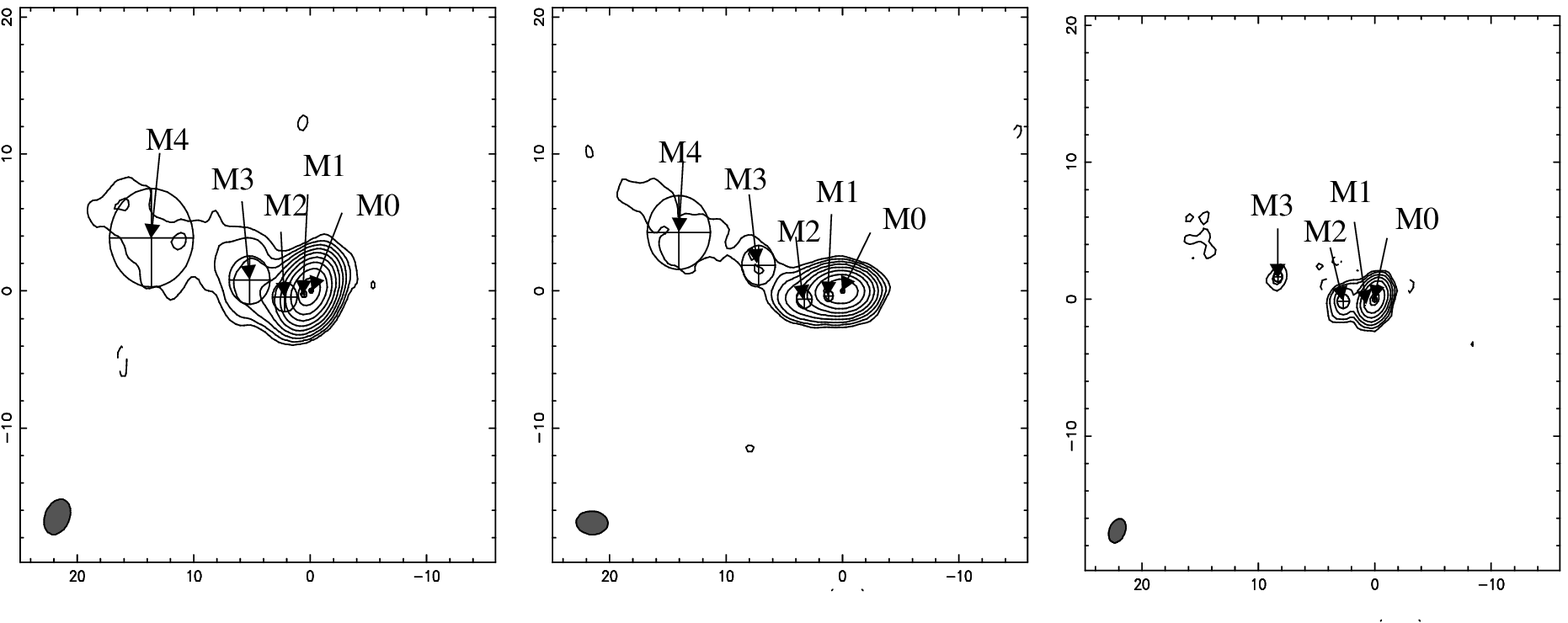}} \caption{ EVN
and VLBA images of 1133+704 at 5 GHz. The epoch of observation from
left to right are:epoch 1997.055, epoch 1998.493, epoch 1999.893.
The axes are labeled in milliarcseconds. Contours are drawn at -1,
1, 2, 4, 8, 16,... times the noise level. Numerical parameters of
the images are given in Table \ref{table1}.}\label{1133map}
\end{figure*}

\begin{figure}
\resizebox{\hsize}{!}{\includegraphics{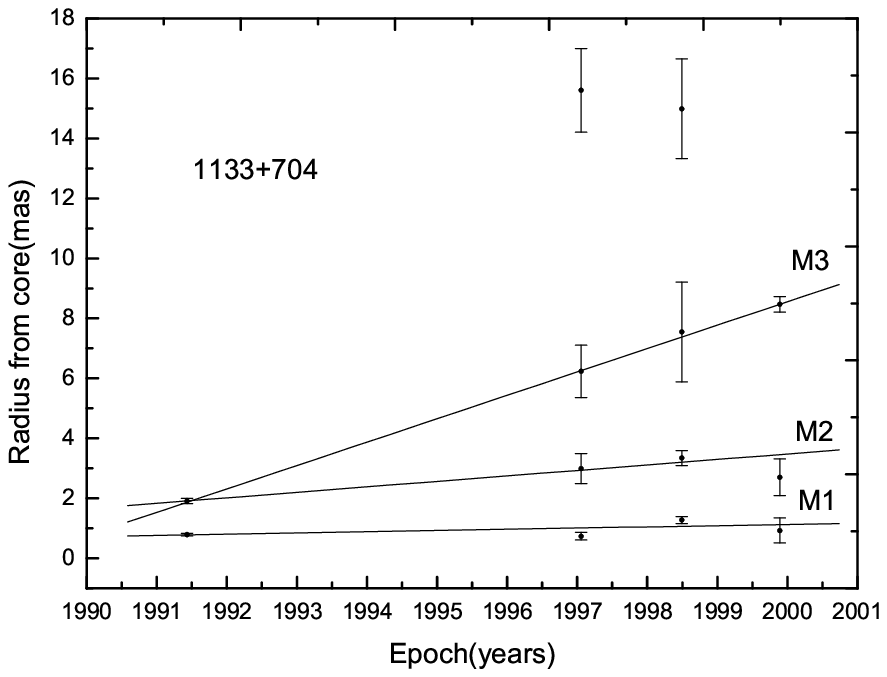}}
\caption{Positions of components with respect to the core at
different epochs from model fitting for 1133+704. The data point at
1991.433 is taken from \cite{koll96a}. All the other epochs are
those presented in this paper, and the lines represent the linear
fitting of the motion for each jet component.} \label{fig:1133}
\end{figure}

{\it 1133+704 --} The images from three epochs (see Fig.
\ref{1133map}) all show that the jet extends at a structural
position angle of $\sim110^{\circ}$, and the outer components extend
to the northeast around $75^{\circ}$. These features are similar to
the parsec scale jet found by \cite{koll96a} and are well-aligned
with the kilo-parsec scale halo and short inner jet presented by
\cite{gir04a} and \cite{gir06}. It can be seen from Fig.
\ref{1133map} that five components are detected in 1133+704, which
is labeled as M0, M1, M2, M3, and M4, respectively. The weighted
linear fits of components M1, M2, and M3 are presented in Fig.
\ref{fig:1133}.
As expected from the fitted line, M3 can be close to M2 at epoch
1991.433; however, it is not detected in the archive data.
As M4 is only detected at two epochs, we do not perform the fit to
this component. It can be seen from Table. \ref{table2} that the
inner components move more slowly, and the outer ones move faster. 

The variability of this source has been investigated on other
wavebands. \cite{xie04} show that the source did not show any
noticeable variability, and it was quiet in the optical band during
their observations. \cite{albert06} have detected very high-energy
$\gamma$-Rays from this source, and an optical outburst in 2006
March was also observed but no evidence of flaring was detected by
the University of Michigan Radio Observatory (UMRAO).

\begin{figure}
\resizebox{3in}{!}{\includegraphics{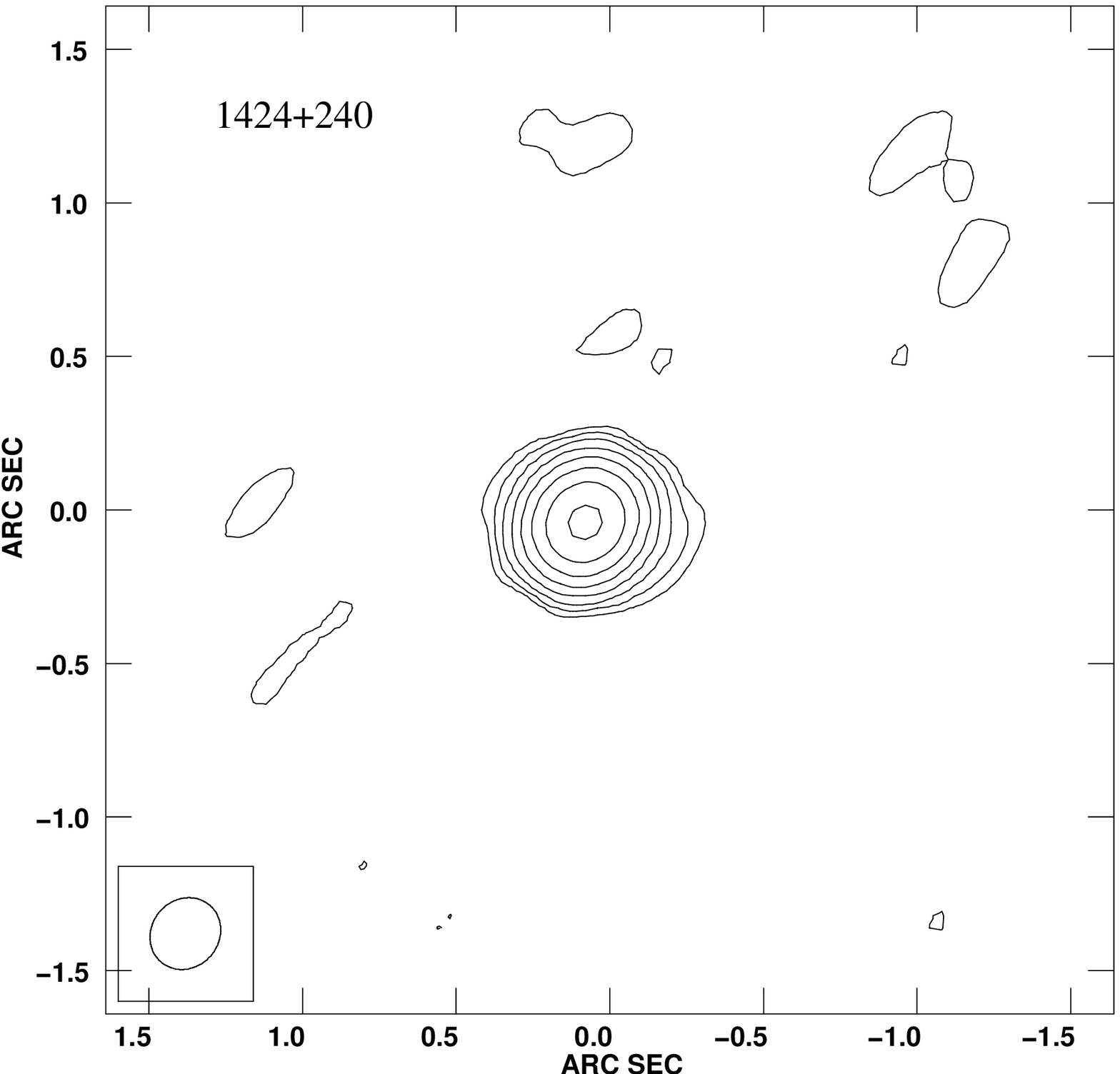}} \caption{The
MERLIN archive image of 1424+240 at 1.6 GHz observed on 1993 Apr 1.
The contour levels are -1, 1, 2, 4, 8, 16, 32, 64... times 0.001745
Jy $beam^{-1}$(3$\sigma$).}\label{1424merlin}
\end{figure}

{\it 1424+240 --} From the images of three epoch VLBI observations
(Fig.\ref{otherall}), the component can not be well-resolved;
however, the jet clearly extends to the southeast with position
angle of about P.A.$=150^{\circ}$. This result is similar with
\cite{fey00}, who modeled this source with two components at 2.3 GHz
and 8.5 GHz, all with jets extending in the nearly same direction.
However, this direction is a little misaligned with the VLA map
found by \cite{rec03}, which shows a compact structure consisting of
a core and either a halo or roughly collinear jets extending north
P.A.$= -10^{\circ}$ and south P.A.$= -175^{\circ}$. The MERLIN
archive data (see Fig. \ref{1424merlin}) shows a very compact
structure, and the jet direction is not evident.

{\it 1542+614 --} There have no previous observations and archive
data for this source. The redshift is unknown. We present our
observations in Fig. \ref{otherall}. The source shows a compact core
and straight jet extending to P.A.$=110^{\circ}$ on a parsec scale.

\begin{figure}
\resizebox{3in}{!}{\includegraphics{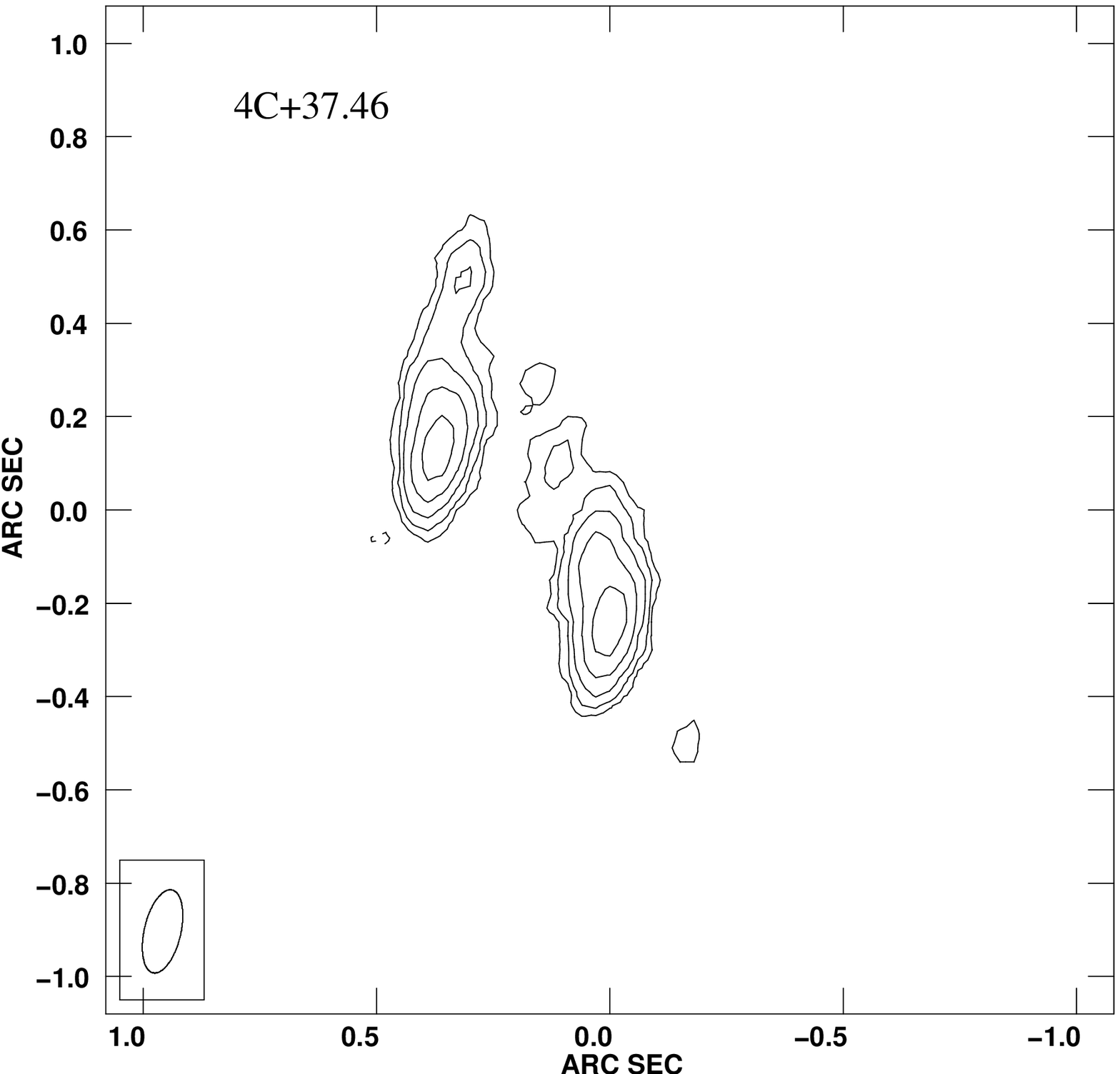}} \caption{The
MERLIN archive image of 4C+37.46 at 5 GHz observed on 1999 May 6.
The contour levels are -1, 1, 2, 4, 8, 16, 32, 64... times 0.001745
Jy $beam^{-1}$(3$\sigma$).}\label{4c3746}
\end{figure}

{\it 4C+37.46 --} We present our image in Fig. \ref{otherall}, in
which the source clearly shows one jet towards the northeast P.A.$=
55^{\circ}$. The MERLIN archive image (see Fig. \ref{4c3746}) also
shows a jet in the same direction. The jet direction is consistent
with \cite{mare06}, in which this source could be core-jet-lobe
source, either with a steep spectrum or some diffuse structure. This
source may also be a well aligned source.

\begin{figure*}
\centering {\includegraphics[width=12cm] {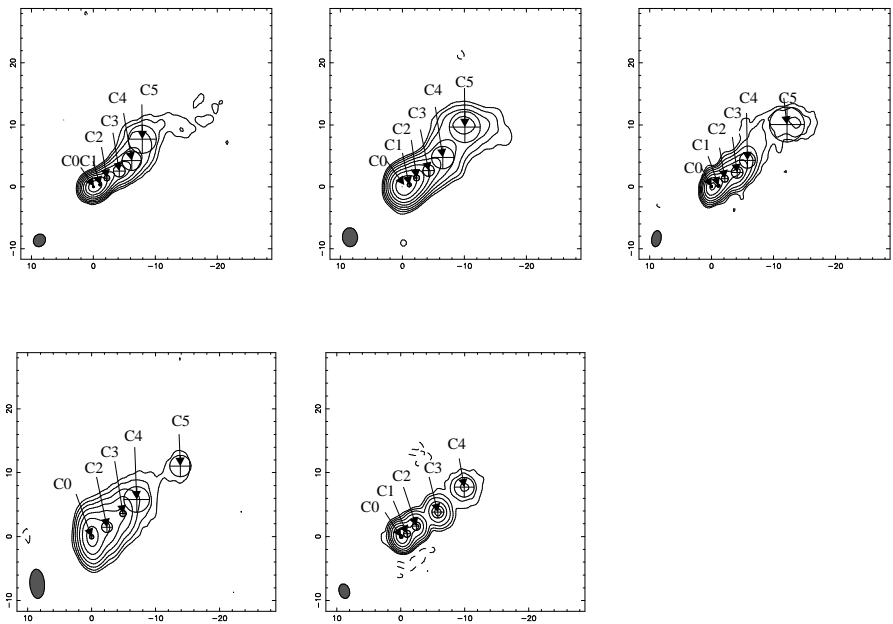}}
\caption{EVN and VLBA images of 1727+502 at 5 GHz. The top from left
to right are : epoch 1995.534, epoch 1997.055, epoch 1998.493. The
bottom from left to right are: epoch 1999.893, epoch 2002.403. The
axes are labeled in milliarcseconds. Contours are drawn at -1, 1, 2,
4, 8, 16,... times the noise level. Numerical parameters of the
images are given in Table \ref{table1}. }\label{1727map}
\end{figure*}

\begin{figure}
\resizebox{\hsize}{!}{\includegraphics{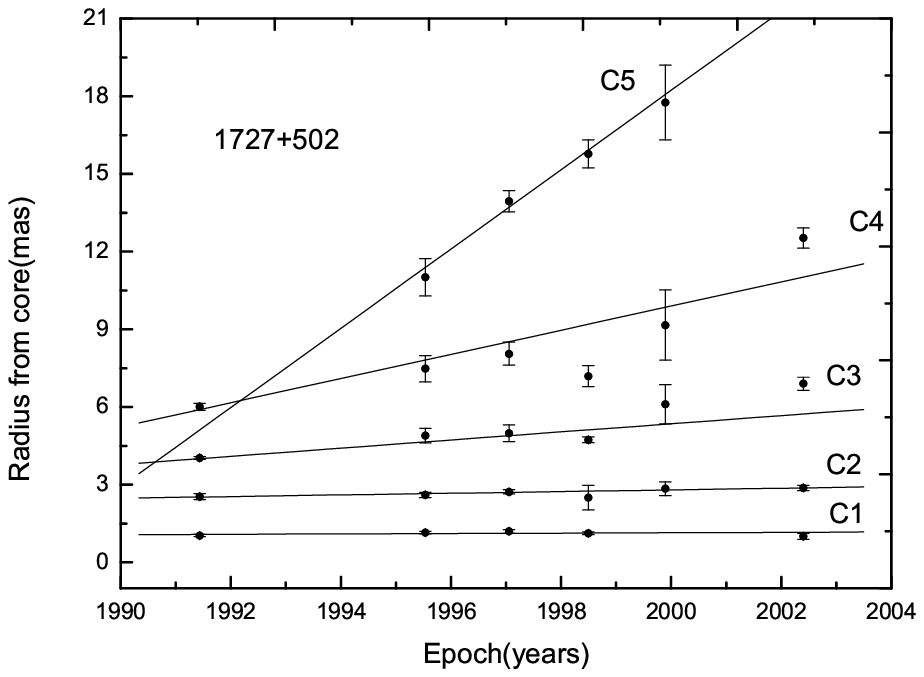}}
\caption{Positions of components with respect to the core at
different epochs from model fitting for 1727+502. The data point at
1991.433 is taken from \cite{koll96a}. All the other epochs are
those presented in this paper, and the lines represent the linear
fitting of the motion for each jet component.} \label{fig:1727}
\end{figure}

{\it 1727+502 --} The large amplitude variability of this source has
been detected on several occasions. A violent variation of 2.1 mag
in the optical band was reported by \cite{scott76}. \cite{fan99}
detected the variations of $\Delta{H}=0.57~\rm mag$ and
$\Delta{K}=0.82$ mag. This source was observed by \cite{koll96a} in
1991 at 6 cm and showed a core-jet morphology. \cite{gir04a} and
\cite{gir06} show that this source displays a remarkable alignment
from the parsec to the kilo-parsec scale. The multi-epoch images in
Fig. \ref{1727map} show a core and a jet extending to northeast with
a position angle of about ~$-55^{\circ}$. The modelfit parameters of
this source were presented in Table \ref{tab:model}. It can be seen
from Fig. \ref{1727map} that six components were detected, which are
labeled with C0, C1, C2, C3, C4, and C5. The weighted linear fit of
components C1, C2, C3, C4, and C5 are shown in Fig. \ref{fig:1727},
and their measured proper motion and apparent speed are listed in
Table \ref{table2}. Interestingly, the component C5 was not detected
at epoch 1991.433 (data from Kollgaard et al. 1996a ) and 2002.403.
From the fitted line for C5, the position of C5 at epoch 1991.433 is
expected to be close to that of component C4; therefore, it is
possible that these components are mixed together, causing
difficulty in detection. At 2002.403, C5 very likely moved far away
from the core and became very weak, so it was not detected. Due to
the poor uv-coverage of our data and the fact that the beam size at
this epoch is larger than other observations, we did not detect
component C1 at 1999.893 (see Fig. \ref{fig:1727}). It is apparent
that C3 and C4 at epoch 1998.493 was closer to the core than the
expected distance from the fit. Actually, it seems that these two
components move slowly (or were even stationary) before 1998.493,
but move quickly afterwards, with the proper motion of C4 comparable
to that of C5. However, the present data do not allow us to
investigate this possibility further. From Table \ref{table2}, we
find that the apparent speed of components increases with the
distance from the core. When looking in detail, components C4 and C5
show superluminal motion, while C1, C2 and C3 show subluminal
motions. C1 is the closest component to the core, and it also shows
the smallest proper motion. Actually, the error in the proper motion
of C1 is larger than its proper motion. Therefore, it is possible
that this component is stationary within the available observations.

\begin{figure*}
\centering {\includegraphics[width=12cm]{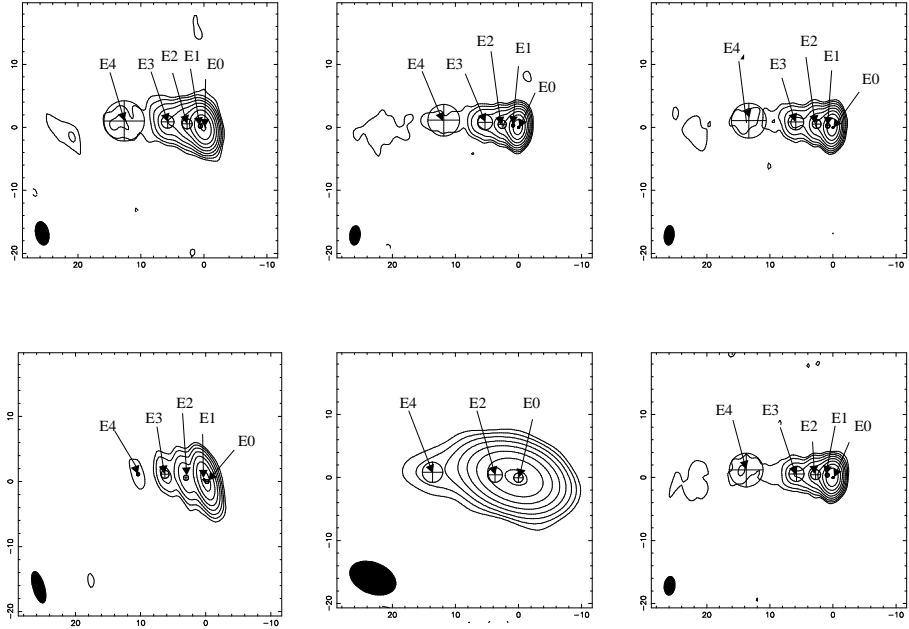}} \caption{EVN
and VLBA images of 1741+196 at 5 GHz. The top from left to right
are:epoch 1997.055, epoch 1997.378, epoch 1998.493. The bottom from
left to right are:epoch 1999.893, epoch 2002.400, epoch 2003.447.
The axes are labeled in milliarcseconds. Contours are drawn at -1,
1, 2, 4, 8, 16,... times the noise level. Numerical parameters of
the images are given in Table \ref{table1}.}\label{1741map}
\end{figure*}

\begin{figure}
\resizebox{\hsize}{!}{\includegraphics{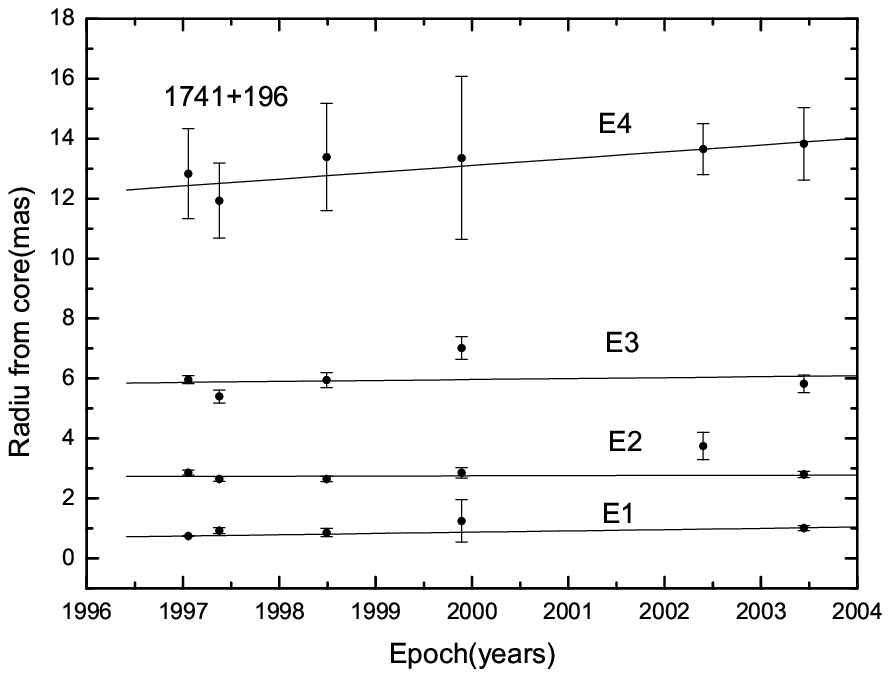}}
\caption{Positions of components with respect to the core at
different epochs from model fitting for 1741+196. All the data
points are presented in this paper, the lines represent the linear
fitting of the motion for each jet component except the data point
at 2002.400, as the baseline of this epoch is much shorter than the
others.} \label{fig:1741}
\end{figure}

{\it 1741+196 --} The six-epoch images of this source in Fig.
\ref{1741map} all show a very straight jet extending to the east
with P.A.$= 80^{\circ}$. This is similar to the result in
\cite{rec03}, in which one epoch VLBA map was presented, and the jet
direction is nearly aligned with the VLA snapshot in \cite{per96}
with
$\Delta{\rm P.A.}$ = $5^{\circ}$. 
Five components are detected in Fig. \ref{1741map}, which are
labeled as E0, E1, E2, E3, and E4. The weighted linear fit of E1,
E2, E3, and E4 are shown in Fig. \ref{fig:1741}, and their measured
proper motion and apparent speed are listed in Table
\ref{table2}. 
When we perform fits to the components, we exclude the data at epoch
2002.400, because the beam size of this epoch is much larger than
others. 
While E4 shows a superluminal motion, E1, E2, E3 all show a rather
small apparent speed. For components E2 and E3, the error in proper
motion is actually larger than their proper motion. In view of this
fact and the rather low value of proper motion, E2 and E3 could be
stationary (within the error) in the time duration covered by the
presenting available observations.

\section{Properties of a sample of BL Lac objects}

\cite{nieppola2006} collected a large amount of multi-frequency data
for the objects in the Mets$\ddot{a}$hovi radio observatory BL Lac
sample. The Mets$\ddot{a}$hovi BL Lac sample includes 381 objects
selected from the Veron-Cetty \& Veron BL Lac Catalogue (Veron-Cetty
\& Veron 2000, hereafter VCV2000), and 17 objects from the
literature, of which many sources are from the well-known BL Lac
samples like 1Jy, S4, S5, Einstein Medium Sensitivity Survey (EMSS),
Einstein Slew Survey, and DXRBS. The authors argue that this sample
is supposed to have no selection criteria (other than declination)
in addition to the ones in the original surveys. Based on the
multi-frequency data, the SED of each source were constructed in the
log $\rm\nu$ - log $\nu F_{\nu}$ representation. The synchrotron
component of the SED was fitted with a parabolic function
\begin{equation}y=Ax^2+Bx+C\end{equation} in order to determine the
synchrotron peak frequency $\nu_{\rm peak}=-B/2A$. The objects were
assigned an LBL/IBL/HBL classification according to $\nu_{\rm
peak}$: for LBLs log\,$\nu_{\rm peak}<$\,14.5, for IBLs
14.5\,$<\,$log\,$\nu_{\rm peak}<$\,16.5, and for HBLs log\,$\nu_{\rm
peak}>$\,16.5.

We collected the available data at 330 MHz, 360 MHz, 408 MHz, and
1.4 GHz from the Astrophysical Catalogues Support System (CATs)
\footnote{http://cats.sao.ru/} maintained by the Special
Astrophysical Observatory, Russia, and also the available VLA or
MERLIN core and extended flux for their sample, resulting in a
sample of 170 BL Lac objects. The redshift is known for most of the
sources from NED and the literature. For the sources without
redshift, we adopt the average redshift of 0.473, 0.302, 0.271 for
LBLs, IBLs, and HBLs, respectively. The sample of BL Lac objects is
listed in Table 4: col. 1 the source IAU name (J2000); col. 2 the
source alias name; col. 3 the redshift; col. 4 the intrinsic
synchrotron peak frequency (see Section 4.4); col. 5 the total 408
MHz radio power; col. 6 the 5 GHz core luminosity; col. 7 the 5 GHz
or 1.4 GHz extended flux density; col. 8 the references; col. 9 the
Doppler factor (see Sect. 4.4); cols. 10 - 12 the viewing angle
estimated assuming $\Gamma=$ 3, 5, and 10, respectively (see Sect
4.5). In the following sections, we investigate the various
properties, e.g. the Doppler factor, viewing angle for this sample.


\subsection{Method of estimating Doppler factor}\label{method}
A reliable determination of the Doppler factor, $\delta$, is a key
step in studying the origins of the physical process in the compact
emission regions of AGNs \citep{dopvar}.
There are several methods available from the literature that can be
used to estimate the Doppler factor. \cite{ghi93} estimated the
Doppler factor for a sample of AGNs, based on the synchrotron
self-Compton model. This method needs the angular size of the core
from high-resolution VLBI observation and assumes that the X-ray
flux originates from the self-Compton components. However, this
method is limited in practice for the case of BL Lac objects. On the
one hand, the detailed VLBI studies on BL Lac objects are only
available for bright objects and a few other outstanding sources
\citep{gir06}. On the other hand, while both the soft and the hard
X-ray bands are dominated by the inverse Compton process in FSRQs,
they are dominated by the synchrotron process in HBL, and the
synchrotron flux dominates in the soft band and the flatter Compton
component appears at higher X-ray energies in LBL \citep{donato01}.

In this work, we use the method suggested by \cite{giova01} and
\cite{gir04a}. Generally, the low-frequency radio power is a
reliable indicator of the intrinsic radio power as it is only
marginally affected by Doppler-boosted compact components
\citep{bondi01}, while the relativistic boosting at the base of the
jet affects the observed radio power of the core. \cite{giovan88}
and \cite{giova01} find a general correlation between the intrinsic
core radio power and total radio power for a sample of radio
galaxies
\begin{equation}{\rm log~ P_{ci5}=~0.62~log~ P_t+8.41}\end{equation}
where P$_{\rm ci5}$ is the intrinsic core 5 GHz radio power derived
assuming $\Gamma= 5$ (see \cite{giova01} for details), and $\rm
P_{t}$ is the total radio power at 408 MHz. From the measured total
408 MHz radio power, we can derive the intrinsic core radio power.
Thus, the Doppler factor can be derived if we know the observed
radio core luminosity. As it is currently believed that most BL Lac
objects are beamed FR I radio galaxies , \cite{gir04a} have applied
this correlation to 30 BL Lac objects and find that the distribution
of either the total radio power or the intrinsic core radio power
derived from this correlation is similar with the FR I sample in
\cite{giova01}. This indicates that Eq. (2) is applicable to our
large sample of BL Lac objects, although this relation was made from
a sample of mainly radio galaxies, and only
few BL Lac objects were included. 

\subsection{Low-frequency radio power}\label{intrinpower}

The anti-correlation between the synchrotron peak frequency and
the intrinsic power was first suggested by \cite{fossati98} and
\cite{ghisel98} for samples of blazars. 
\cite{nieppola2006} find an anti-correlation between the luminosity
at 5 GHz and 37 GHz with $\nu_{\rm peak}$ for a large sample of BL
Lac objects. However, the 5 GHz and 37 GHz luminosity can be
severely affected by the beaming effect from the radio jets and
can't be used to indicate of the intrinsic radio power.
\cite{bondi01} and \cite{gir04a} suggest that the low-frequency
radio luminosity of BL Lac objects can indicate the intrinsic power
from the radio jet. However, most of the objects in their works are
HBLs and IBLs. For LBLs, especially those in the 1Jy sample, as
there is a selection criterion of the radio spectral index being
less than 0.5 ($\alpha_r\leq0.5$; $S_\nu \propto\nu^{-\alpha_r}$),
the low-frequency radio emission may still be Doppler boosted
\citep{cao03}. Indeed, \cite{liu06} find that the low-frequency
radio emission of sources with a flat spectrum are likely to be
Doppler-boosted, although only radio-loud quasars are considered in
their work.

%

\begin{figure}
\resizebox{\hsize}{!}{\includegraphics{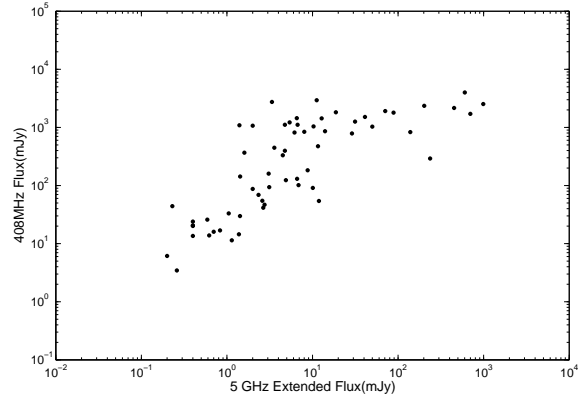}} \caption{The
relationship between the radio 5 GHz extended flux and the radio
low-frequency 408 MHz flux. } \label{fig:flux}
\end{figure}

To test whether the low-frequency radio emission is Doppler-boosted,
we selected 61 BL Lac objects from \cite{nieppola2006}, which have
VLA observations and extended flux. These sources are listed in
Table \ref{tab:core}, including all powerful sources of the 1 Jy
sample, and some other type of BL Lac objects. After extrapolating
from the 1.4 GHz extended flux using $\alpha=1.0$ for 15 sources
without 5 GHz extended flux,  we plot the 5 GHz extended flux with
the total 408 MHz flux in Fig. \ref{fig:flux}, which is directly
from the available measurements, or extrapolated from 330 MHz, 360
MHz, or the total 1.4 GHz radio flux (see below). We find a
generally good correlation. However, the scatter is
obviously seen. 
This can be caused by the difference in the spectral index of two
bands between sources, while both emissions are not influenced by
the beaming effect. However, the possibility that the low-frequency
radio emission in some sources is indeed Doppler boosted can't be
excluded completely with present data. Nevertheless, we believe that
the low-frequency radio emission is intrinsic, as most of the
sources follow the correlation.

\begin{figure}
\resizebox{\hsize}{!}{\includegraphics{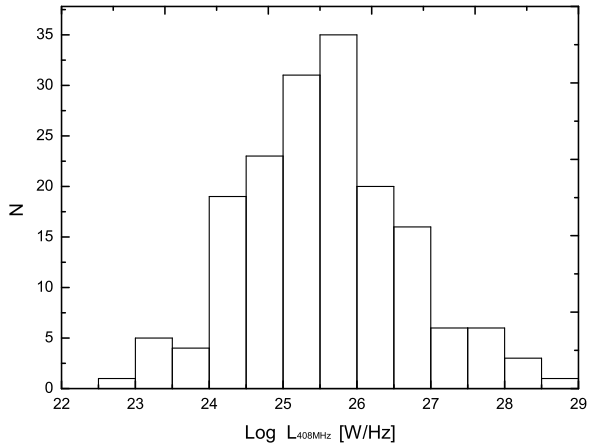}} \caption{The
distribution of 408 MHz luminosity for BL Lac objects selected from
\cite{nieppola2006}.} \label{fig:hist}
\end{figure}

To calculate the Doppler factor using the method in Sect. 4.2, we
had to obtain the total 408 MHz luminosity. Unfortunately, not all
sources in our sample had an observed 408 MHz flux. We adopted three
methods of calculating 408 MHz luminosity according to the available
data. For BL lac objects with the extended flux available, we
estimated the 408 MHz luminosity by extrapolating from the 5 GHz or
1.4 GHz extended flux assuming a spectral index of $\alpha=1.0$. For
the sources with observed low-frequency radio flux, we estimated the
total 408 MHz luminosity after a K-correction using spectral index
$\alpha=1.0$. For the remaining sources, neither the extended flux
nor the low-frequency flux was available, so we extrapolated from
the total 1.4 GHz flux density with a spectral index of
$\alpha=0.207$, which is the average spectral index between 408 MHz
and 1.4 GHz of the BL Lac objects in our sample.
The estimated total 408 MHz radio power is listed in Col. 5 of Table
\ref{tab:core}. To compare with the FR I radio galaxies sample of
\cite{zirbel}, we converted the total 408 MHz radio luminosity into
the same cosmological frame ($H_{0}=50 \rm {~km ~s^ {-1}~Mpc^{-1}}$
and $q_{0}=0$), as used in their calculation. The distribution of
this total 408 MHz radio power is shown in Fig. \ref{fig:hist}.
Although our sample contains about two thirds of the BL Lac objects
from the 1Jy sample, we find that the mean and median values of the
408 MHz radio power ( mean$\rm =25.56~W~Hz^{-1}$ and median$\rm
=25.52~W~Hz^{-1}$) are consistent with those of the low-luminosity
FR I radio galaxies in \cite{zirbel}, which is mean$\rm
=25.50\pm0.12~W~Hz^{-1}$ and median$\rm =25.38~W~Hz^{-1}$. This, on
one hand, supports the scenario that FR I radio galaxies are the
parent population of BL Lac objects; and on the other hand, it
indicates that the estimated total 408 MHz radio power of our BL Lac
objects is intrinsic, as it is comparable to FR I radio galaxies.

\subsection{Doppler factor of BL Lac objects}
\label{doppler} According to the unified schemes, classes of
apparently different AGN might actually be intrinsically similar,
when only viewed at different angles with respect to the line of
sight, it was suggested that BL Lac objects are FR I radio sources
viewed at relatively small angles to the line of sight and that the
relativistic beaming has an enormous effect on the observed
luminosities, so determining the Doppler factor is important in
studying the properties of BL Lac objects, especially the
differences between different types of BL Lac objects. To estimate
the Doppler factor using the method described in Sect. \ref{method},
we collected 5 GHz VLA or MERLIN core flux of 170 BL Lac objects
(see Table \ref{tab:core}). The 5 GHz core luminosity was calculated
with a K-correction assuming a spectral index of $\alpha$=0.
The Doppler factor is then estimated by using the 5 GHz core
luminosity, the estimated radio power at 408 MHz (see Sect.4.3), and
Eq. (2). In this work, we adopt $\rm P_{c,o}= P_{c,i}\delta^{2+
\alpha}$ (corresponding to a continuous jet), assuming $\alpha=0$,
where $\rm P_{c,o}$ is the observed 5 GHz core luminosity and $\rm
P_{c,i}$ the intrinsic one calculated from Eq. (2).

\begin{figure}
\resizebox{\hsize}{!}{\includegraphics{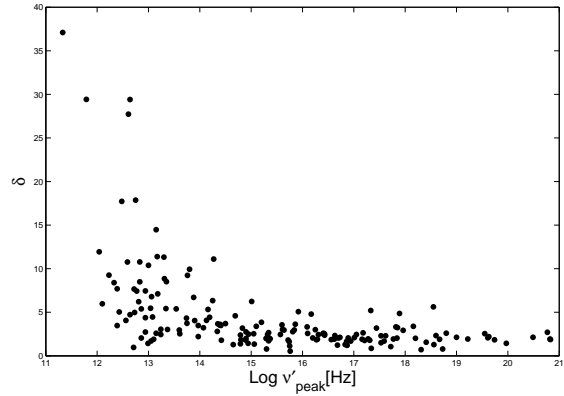}} \caption{The
Doppler factor versus the intrinsic synchrotron peak frequency
$\nu_{\rm peak}^{'}$ (see text for details). } \label{fig:dop}
\end{figure}

The synchrotron peak frequency $\rm \nu_{peak}\propto \it B \rm
\delta \gamma_{peak}^{2}$, therefore it may not be intrinsic since
the Doppler factor can vary between sources in our sample. Once the
Doppler factor can be estimated, it would be fundamental in
calculating the intrinsic peak frequency. In this work, we calculate
the intrinsic synchrotron peak frequency $\nu^{'}_{\rm
peak}$=$\nu_{\rm peak}(1+z)/\delta$, where $\nu_{\rm peak}$ is from
\cite{nieppola2006}. The factor (1+$z$) is included since the
$\nu_{\rm peak}$ of \cite{nieppola2006} is determined from the log
$\nu$-log $\nu f_{\nu}$ panel in the observer's frame.
The distribution of the Doppler factor and the intrinsic peak
frequency is shown in Fig. \ref{fig:dop}.
While all IBLs and HBLs have a Doppler factor below 10, LBLs have a
wide spread for the Doppler factor, with most sources below 20, and
it can be over 30 in a few cases. It seems that the Doppler factor
systematically decreases with increasing peak frequency; however, it
is similar for IBLs and HBLs. Indeed, the Spearman rank correlation
analysis shows an anti-correlation between the Doppler factor and
$\nu^{'}_{\rm peak}$ at $\gg99.99$ per cent significance level. This
may indicate that the Doppler boosting is systematically more
pronounced in LBLs than in IBLs and HBLs, although a fraction of
LBLs are comparable to IBLs and HBLs.

\begin{figure*}
\centering {\includegraphics[width=12cm]{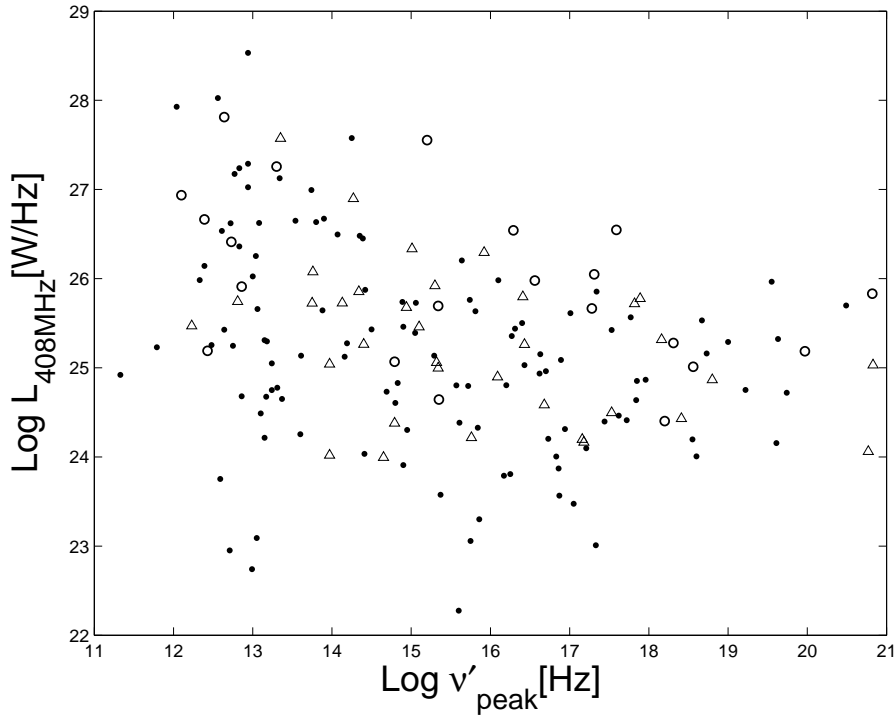}}
\caption{The low-frequency 408 MHz luminosity versus the intrinsic
synchrotron-peak frequency $\nu_{\rm peak}^{'}$. The open circles
are the sources with the direct measurements of 408 MHz radio flux
and known redshift, the open triangles are the sources calculated
from either the 360 MHz, or 330 MHz radio flux with known redshift,
while the solid circles are those estimated from the extended flux
(at 5 GHz or 1.4 GHz), the total 1.4 GHz radio flux, or
low-frequency radio flux (with unknown redshift) (see text for
details).} \label{fig:totalc}
\end{figure*}

The estimation of Doppler factor enables us to investigate the
relationship between the intrinsic radio power and the intrinsic
synchrotron peak frequency. The relationship between the intrinsic
peak frequency and 408 MHz radio power is shown in Fig.
\ref{fig:totalc}. It can be seen that LBLs are statistically more
powerful than other BL Lac objects. There is a significant
anti-correlation between luminosity and $\nu_{\rm peak}^{'}$ at
$\gg99.99\%$ confidence level from the Spearman Rank Correlation
analysis.
Although we used the low-frequency radio emission to indicate the
intrinsic radio power and correct the peak frequency for Doppler
factor, our results in Fig. \ref{fig:totalc} are very similar to
those of radio power at 5 GHz in \cite{nieppola2006}. Several
low-luminosity LBLs are apparently seen, and they even reach lower
luminosities than any of the HBLs.

We note that the 408 MHz radio power in our sample is calculated in
inhomogenous ways, and the redshift of some sources is actually
unknown and tentatively taken as the average value of subclasses. To
verify how the correlation depends on the data, we check with a
sample that has the measured 408 MHz flux and known redshift. Still,
these twenty-one sources ( Fig. \ref{fig:totalc}) show a strong
anti-correlation at about a 98 per cent confidence level. However,
due to the small number of these sources, we add the sources with
known redshift and 408 MHz radio power extrapolated from either 330
MHz or 360 MHz radio flux ( Fig. \ref{fig:totalc}), resulting in a
sample of 56 sources, of which the uncertainty from extrapolation is
minimized. It is clearly seen that these sources follow the general
trend of whole sample. The Spearman rank correlation analysis shows
a significant anti-correlation at about a 99.9 per cent confidence
level. We are thus confident that the inhomogenous data derivation
does not influence our results.

Despite much more scatter in our figure than in the corresponding
one of \cite{fossati98}, the anti-correlation between luminosity and
$\nu_{\rm peak}^{'}$ is still significant and possibly caused by the
intrinsic physical differences (such as magnetic fields, electron
energies), as suggested by \cite{sambr96}. As jet formation is
closely connected with the central engine \citep{liu06}, this
correlation can also be caused by the common relationship between
both the radio luminosity
and peak frequency and the central engine, as suggested by \cite{jianmin02} 
that the peak frequency is significantly correlated with the
accretion rate.
Although our sample is relatively large, still only a small fraction
of low-luminosity LBLs are included in the sample. As a matter of
fact, there is no evidence of very high-luminosity HBLs in the
sample of \cite{nieppola2006}, which is  our parent sample.

\subsection{Viewing angle}
\begin{figure}
\resizebox{\hsize}{!}{\includegraphics{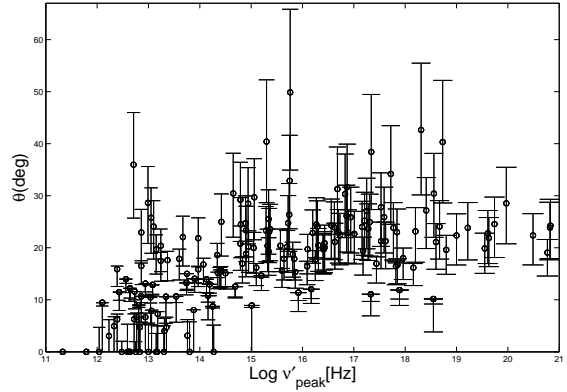}} \caption{The
viewing angle versus the intrinsic synchrotron peak frequency
$\nu_{\rm peak}^{'}$. The viewing angle is estimated by assuming a
Lorentz factor of 5 and using the estimated Doppler factor in Sect.
4.4. The errorbar is indicated with the viewing angle of Lorentz
factors of 3 and 10.} \label{fig:angle}
\end{figure}

There have been several models trying to explain the differences in
different types of BL Lac objects. It was originally explained as
the orientation effects, with HBLs being observed at larger
inclination to the jet axis \citep{urry95}. However, \cite{sambr96}
argue that it is difficult to model the detailed transition from an
HBL to an LBL SED only in terms of orientation only and instead
suggest a continuous change in the physical parameters of the jet. 
To study the viewing angle distribution for our sample of BL Lac
objects, we estimated the viewing angle for each object assuming a
Lorentz factor of $\Gamma$= 5, in combination with the estimated
Doppler factor in Sect. \ref{doppler}. The difference between this
angle and that of $\Gamma$= 3 and $\Gamma$= 10 are taken as the
uncertainty. The distribution of the viewing angle and intrinsic
synchrotron peak frequency $\nu_{\rm peak}^{'}=\nu_{\rm
peak}(1+z)/\delta$ is shown in Fig. \ref{fig:angle}, in which the
errorbar are indicated with the viewing angle from $\Gamma$= 3 and
$\Gamma$= 10. It can be seen that most of LBLs are in the range of
angles ($0^{\circ}$, $20^{\circ}$), while most of IBLs and HBLs are
in ($15^{\circ}$, $30^{\circ}$). As an intermediate subclass between
the LBLs and HBLs, the IBLs' viewing angles are larger than that of
LBLs; however, it is indistinguishable from those of HBLs. Moreover,
BL Lac objects, as a whole, mostly have viewing angle smaller than
$30^{\circ}$. This is consistent with the unified scheme in which BL
Lac objects are the aligned version of FR I radio galaxies (Urry \&
Padovani 1995). It clearly shows that there is a significant
positive correlation between viewing angle and $\nu_{\rm peak}^{'}$
at $\gg$99.99 per cent significance level. Actually, this can be
expected simply from the relationship between the Doppler factor and
$\nu_{\rm peak}^{'}$, if an uniform Lorentz factor is adopted for
all sources. Therefore, it is not surprising that LBLs have smaller
viewing angles compared to IBLs and HBLs, since their Doppler factor
is systematically larger.

Usually, the viewing angles can be estimated by using the Doppler
factor and the measurements of proper motion, however, the latter
are not available for most of sources. Using the measured proper
motion and the Doppler factor, we calculated the Lorentz factor for
the three sources we observed. We find that the estimated Lorentz
factor is generally consistent with the assumption of $\Gamma= 5$,
and the range of $\Gamma$ between 3 and 10 (see Table 3).
This indicates that the angle got by assuming a Lorentz factor of 5
can be reasonable for our sample. Certainly, the adoption of a
Lorentz factor of 5 only makes sense for our statistical sample
investigation and can't be used for an individual source. We believe
that the exact value of Lorentz factor will not change our results
significantly. From Fig. \ref{fig:angle}, it seems that the BL Lac
objects as a single population possess a continuous range of viewing
angle and a positive correlation between the viewing angle and the
$\nu_{\rm peak}^{'}$. Similar to the Doppler factor distribution,
the viewing angle increases with the increasing peak frequency from
LBLs to IBLs, but, it smooths out upward.

\section{Discussion }

In sect. \ref{comment}, our observations have shown that all the
jets of seven BL Lac objects are very straight, and they do not
exhibit large bending between the small and large scales. This
result is similar with the observations of HBLs in \cite{rec03}, in
which two of our sources are included. This indicates that these
sources are either intrinsically straighter or are seen further
off-axis than LBL jets. In three sources with measured proper motion
(Sect. \ref{comment}), it shows that different components of the
same source have different apparent speeds, and the apparent speed
increases with the distance from the core. 
 This is similar to the results in \cite{pin06}, in which BL
Lac object 0235+164 and two FSRQs were studied. This phenomena will
need a change in either bulk Lorentz factor or viewing angle.
However, these changes would also yield predictable changes in the
brightness of components due to changing Doppler boosting, and
these changes are not seen. 

While recent discoveries of low-luminosity LBLs and possible
high-luminosity HBLs (Padovani et al. 2003; Caccianiga \& March\~{a}
2004; Giommi et al. 2005) seems to undermine the blazar sequence,
because an anti-correlation between radio power and peak frequency
was found for a large sample of BL Lac objects \citep{nieppola2006},
though with a large scatter. 
We improved their work by using the intrinsic low-frequency radio
power and the intrinsic synchrotron peak frequency, instead.
It appears that LBLs are intrinsically more powerful than other
types of BL Lac objects, and the distribution of the low-frequency
power of BL Lac objects as a whole is comparable to those of FR I
radio galaxies. Although there is a correlation between the
intrinsic low-frequency power and intrinsic peak frequency, the
scatter of this relation seems large, which contrast with the much
tighter relation of `blazar sequence'. More important, some LBLs
have low radio power, which is difficult to explain by the radiation
cooling process, which was used to explain the blazar sequence. In
fact, these sources tend to weaken the blazar sequence; however,
they are not enough to eliminate the correlation in this work.
While the blazar sequence was suggested for a sample of blazars
including FSRQs, of which the peak frequency can be lower than those
of LBLs, our sample only includes BL Lac objects, although our
sample is much larger. Moreover, the original peak frequency used in
this work is directly adopted from \cite{nieppola2006}, which was
derived by simply fitting the synchrotron component of the SED with
a parabolic function. The determination of peak frequency is thus
influenced by the data quality and sampling. As \cite{nieppola2006}
discuss, the peak frequencies of the most extreme objects can be
exaggerated by using a simple parabolic function in the fitting
procedure. Furthermore, the large scatter in Fig. \ref{fig:totalc}
implies that other parameters may be at work, and the scatter can be
much lower once this parameter is included. Besides the significant
anti-correlation between the intrinsic low-frequency luminosity and
intrinsic peak frequency, there is a positive correlation between
the viewing angle and intrinsic peak frequency. LBLs apparently have
smaller viewing angle, while the viewing angle of IBLs is comparable
to that of HBLs. We conclude that the intrinsic synchrotron peak
frequency is related not only to the intrinsic luminosity, but maybe
also to the viewing angle. In fact, some HBLs like Mrk 421 and Mrk
501 are supposed to have $\delta$$\geq$10, which was estimated by
\cite{celot98} based on their TeV flux variability. However, the
studies of jet morphology, core dominance, the fit to the trend of
the jet brightness, and Full-Width Half-Maximum (FWHM) all show that
the radio jet emission is oriented to larger angles with respect to
the line of sight (Giroletti et al. 2004b; Giroletti et al. 2006).

There are only a small fraction of low-luminosity LBLs in our
sample, and actually no evidence of very high-luminosity HBLs
\citep{nieppola2006}. This precludes us from draws any firm
conclusion about the validity of the blazar sequence. Actually, the
FSRQs with a high $\nu_{\rm peak}$ found in DXRBS do not reach the
extreme $\nu_{\rm peak}$ values of HBLs. It is still unclear whether
this fact indicates an intrinsic, physical limit to this parameter.
An alternative scenario is one where really
high-power-high-$\nu_{\rm peak}$ blazars have their thermal emission
swamped by the non-thermal, featureless jet emission, which makes
their redshift
determination impossible. 
While our results seem to undermine the blazar sequence, the
anti-correlation between the intrinsic emission and intrinsic peak
frequency needs further investigation. It is important to compile a
larger sample including various subclasses of blazars, especially
blue quasars, low-luminosity LBLs, and high-luminosity HBLs and to
calculate the accurate peak frequency and intrinsic luminosity. Only
then, the physical mechanism of the different peak frequencies can
be explored. While the peak frequency is believed to be determined
by the magnetic strength and the energy distribution of relativistic
electrons,
the shift of the peak frequency in blazars can give us clues to this
information.
It might also be important to investigate the relationship between
the SEDs of blazars and the central engine, such as accretion rate
\citep{jianmin02} and/or black hole mass.

\section{Conclusions}
In this paper, we have presented EVN and VLBA images at 5 GHz for
seven BL Lac objects selected from an RGB sample \citep{laure99}. We
collected multi-epoch data from EVN and the VLBA archive for three
of them. The radio data were collected for a sample of 170 BL Lac
objects selected from \cite{nieppola2006} to study the relation of
the intrinsic synchrotron peak frequency with the intrinsic
low-frequency radio luminosity and the viewing angle. The main
conclusions can be summarized as follows:

\noindent {\it 1.} We found that all seven sources show a core-jet
structure on a pc scale, and no counter-jets
were found. 
The jets of these sources are straight, and only small changes in
the position angle are found between small and large scales. The
superluminal motion is detected in all three sources with the
available multi-epoch data, and the apparent speed of components are
increases with the increasing distance of components to the core.

\noindent {\it 2.} Based on the method suggested in \cite{giova01}
and \cite{gir04a}, we calculated the Doppler factor for 170 BL Lac
objects selected from \cite{nieppola2006}. Using the total 408 MHz
luminosity to indicate the intrinsic radio power and correcting the
synchrotron peak frequency for Doppler factor, we found a
significant anti-correlation between the total 408 MHz luminosity
and the intrinsic synchrotron peak frequency. However, the scatter
is much larger than that of the blazar sequence. Especially, several
low-power LBLs are clearly present. Our results seem to undermine
the blazar sequence.

\noindent {\it 3.} The viewing angle of BL Lac objects are
constrained using the estimated Doppler factor, and assuming a
Lorentz factor of 5. We found a strong positive correlation between
the viewing angle and the intrinsic synchrotron peak frequency. The
BL Lacs show a continuous distribution of the viewing angle, and
LBLs have lower values than for IBLs and HBLs. However, the IBLs are
similar to HBLs for the viewing angle.

In conclusion, our results show that the intrinsic peak frequency is
related not only to the intrinsic radio power (though much looser
than the blazar sequence), but also to the viewing angle. A
complete, large sample of BL Lacs (or blazars) is needed for future
investigations.



\begin{acknowledgements}
We thank the anonymous referee for insightful comments and
constructive suggestions. This work is supported by the NSFC under
grants 10373019, 10333020, and 10543002. The European VLBI Network
is a joint facility of European, Chinese, and other radio astronomy
institutes funded by their national research councils. The National
Radio Astronomy Observatory is operated by Associated Universities,
Inc., under cooperative agreement with the National Science
Foundation. MERLIN is a National Facility operated by the University
of Manchester at Jodrell Bank Observatory on behalf of PPARC. This
research made use of the NASA/ IPAC Extragalactic Database (NED),
which is operated by the Jet Propulsion Laboratory, California
Institute of Technology, under contract with the National
Aeronautics and Space Administration. This work also made use of
Astrophysical Catalogues Support System (CATS) maintained by the
Special Astrophysical Observatory, Russia.
\end{acknowledgements}

\Online

\begin{table*}
 \caption{The VLBI observational log.}
\label{table1}
\begin{tabular}{lllllll}
\hline\hline

Object &   $z$ &Epoch &HPBW & Noise ($3\sigma$) & Peak & Array\\
       &    &      &(mas $\times$ mas,$^\circ$)&(mJy beam$^{-1}$)&(mJy beam$^{-1}$)
&

\\\hline
1011$+$496&   0.2 &  1997.055  &  $2.5 \times 2.08,\, -8.36$   &  0.60  &116.0      &  1     \\
          &       &  1998.493  &  $2.47 \times 1.72,\,-2.76$   &  0.56  &103.0      &  1         \\
          &       &  1999.893  &  $2.8 \times 1.35,\, -38.6$   &  1.83   &81.7       &  2         \\
1133+704  & 0.046 &  1997.055  &  $2.77 \times 2.04,\, -32.5$  &  0.30  &118.0      &  1    \\
          &       &  1998.493  &  $2.72 \times 1.7, \,87.7$    &  0.39  &98.7       &  1         \\
          &       &  1999.893  &  $1.87 \times 1.32, \,-30.4$  &  1.01   &77.5       &  2         \\
1424$+$240&0.16   &  1999.696   &  $2.96 \times 1.92,\, -0.412$ &  0.94  &125.0      &  1     \\
          &       &  1999.893  &  $5.95 \times 1.98,\, 12.8$   &  1.57   &134.0      &  2          \\
          &       &  2003.447  &  $3.02 \times 1.61,\, -8.25$  &  0.60  &164        &  1         \\
1542$+$614& ...   &  1999.893  &  $2.26 \times 1.48,\, 7.24$   &  0.87   &93.4       &  2          \\
4C$+$37.46& 1.271 &  1999.893  &  $4.65 \times 2.17, \,15.7$   &  1.38   &32.2       &  2      \\
1727$+$502& 0.055 &  1995.534  &  $2.12\times 1.87, \,-34.6$   &  0.37   &80.1       &  1     \\
          &       &  1997.055  &  $3.05 \times 2.43, \,3.03$   &  0.35  &85.7       &  1         \\
          &       &  1998.493  &  $2.62 \times 1.52, \,-10.8$  &  0.46  &66.9       &  1         \\
          &       &  1999.893  &  $4.58 \times 2.3, \,4.5$     &  0.93  &81.4       &  2          \\
          &       &  2002.403  &  $2.31 \times 1.64, \,16$     &  0.63  &72.9       &  2          \\
1741$+$196& 0.083 &  1997.055  &  $3.75 \times 2.13, \,11.8$   &  0.37  &103.0      &  1     \\
          &       &  1997.378  &  $3.07 \times 1.67, \,-5.62$  &  0.42  &80.1       &  1          \\
          &       &  1998.493  &  $3.09 \times 1.58, \,-5.37$  &  0.39  &85.4       &  1         \\
          &       &  1999.893  &  $5 \times 1.79, \,16.2$      &  1.14   &98.4       &  2         \\
          &       &  2002.400  &  $7.68 \times 4.92, \,67.8$   &  0.90  &131.0      &  2          \\
          &       &  2003.447  &  $2.98 \times 1.72, \,-6.07$  &  0.41  &100.0      &  1         \\

\hline\hline
\end{tabular}
\vskip 0.1 true cm \noindent 1: VLBA, 2: EVN.
\end{table*}
\begin{longtable}{ccccccccccc}
\caption{\label{tab:model} Model parameters.}\\

\toprule
Source & Epoch &  Component & $S$ (Jy)& $r$ (mas)&P.A. (deg)& $a$ (mas) \\
\midrule \midrule
\endfirsthead  
\midrule

Source & Epoch &  Component & $S$ (Jy)& $r$ (mas)&P.A. (deg)& $a$ (mas) \\
\midrule
\endhead 

\midrule
\multicolumn{4}{r}{Continued\dots} \\
\endfoot 
\endlastfoot
1727+502 ...... & 1995.534   &     C0  &     0.068    &      0.00            &      0.0    &      0.27    \\
& 1995.534   &     C1  &     0.027    &      1.15$\pm$0.05   &    -73.3    &      0.36    \\
& 1995.534   &     C2  &     0.016    &      2.61$\pm$0.11   &    -57.4    &      0.91    \\
& 1995.534   &     C3  &     0.016    &      4.90$\pm$0.28   &    -58.8    &      1.91    \\
& 1995.534   &     C4  &     0.013    &      7.48$\pm$0.51   &    -55.8    &      3.12    \\
& 1995.534   &     C5  &     0.012    &     11.01$\pm$0.72   &    -45.9    &      4.53    \\
& 1997.055   &     C0  &     0.065    &      0.00            &      0.0    &      0.00    \\
& 1997.055   &     C1  &     0.030    &      1.20$\pm$0.71   &    -75.4    &      0.60    \\
& 1997.055   &     C2  &     0.030    &      2.73$\pm$0.08   &    -58.8    &      0.97     \\
& 1997.055   &     C3  &     0.010    &      4.99$\pm$0.32   &    -58.8    &      1.84    \\
& 1997.055   &     C4  &     0.012    &      8.06$\pm$0.44   &    -54.4    &      3.62    \\
& 1997.055   &     C5  &     0.014    &     13.94$\pm$0.41   &    -46.4    &      5.00    \\
& 1998.493   &     C0  &     0.062    &      0.00            &      0.0    &      0.33    \\
& 1998.493   &     C1  &     0.023    &      1.13$\pm$0.06   &    -81.7    &      0.46    \\
& 1998.493   &     C2  &     0.022    &      2.50$\pm$0.48   &    -59.0    &      1.12    \\
& 1998.493   &     C3  &     0.028    &      4.73$\pm$0.11   &    -60.9    &      1.88    \\
& 1998.493   &     C4  &     0.012    &      7.19$\pm$0.40    &    -53.7    &      2.55    \\
& 1998.493   &     C5  &     0.015    &     15.78$\pm$0.54   &    -50.5    &      5.62     \\
& 1999.893   &     C0  &     0.082    &      0.00            &       0.0   &      0.67        \\
& 1999.893   &     C2  &     0.039    &      2.84$\pm$0.26   &     -57.5   &      1.69        \\
& 1999.893   &     C3  &     0.014    &      6.11$\pm$0.76   &     -53.4   &      1.00        \\
& 1999.893   &     C4  &     0.009    &      9.16$\pm$1.35   &     -50.2   &      4.00        \\
& 1999.893   &     C5  &     0.003    &     17.76$\pm$1.44   &     -51.4   &      3.32        \\
& 2002.403   &     C0  &     0.071    &      0.00            &      0.0    &      0.53      \\
& 2002.403   &     C1  &     0.017    &      1.01$\pm$0.11   &    -63.9    &      1.09      \\
& 2002.403   &     C2  &     0.030    &      2.88$\pm$0.10    &    -55.3    &      1.30     \\
& 2002.403   &     C3  &     0.021    &      6.90$\pm$0.25   &    -56.1    &      1.87      \\
& 2002.403   &     C4  &     0.009    &     12.52$\pm$0.39   & -51.9 & 3.091       \\
1741+196 ......& 1997.055   &     E0  &     0.070    &      0.00            &      0.0    &      0.00       \\
& 1997.055   &     E1  &     0.040    &      0.74$\pm$0.013  &     70.2    &      0.65       \\
& 1997.055   &     E2  &     0.018    &      2.86$\pm$0.079  &     77.9    &      1.60       \\
& 1997.055   &     E3  &     0.009    &      5.96$\pm$0.13   &     81.1    &      2.01       \\
& 1997.055   &     E4  &     0.004    &     12.83$\pm$1.50   &     85.2    &      6.41      \\
& 1997.378   &     E0  &     0.073    &      0.00            &      0.0    &      0.35       \\
& 1997.378   &     E1  &     0.019    &      0.93$\pm$0.10   &     70.4    &      0.51       \\
& 1997.378   &     E2  &     0.015    &      2.65$\pm$0.079  &     79.6    &      1.32       \\
& 1997.378   &     E3  &     0.010    &      5.40$\pm$0.22   &     81.3    &      2.41       \\
& 1997.378   &     E4  &     0.004    &     11.93$\pm$1.25   &     84.5    &      5.04       \\
& 1998.493   &     E0  &     0.076    &      0.00            &      0.0    &      0.28       \\
& 1998.493   &     E1  &     0.022    &      0.86$\pm$0.14   &     72.0    &      0.69        \\
& 1998.493   &     E2  &     0.017    &      2.65$\pm$0.090  &     78.0    &      1.41        \\
& 1998.493   &     E3  &     0.009    &      5.95$\pm$0.25   &     81.6    &      2.47        \\
& 1998.493   &     E4  &     0.004    &     13.39$\pm$1.79   &      85.4   &      5.55        \\
& 1999.893   &     E0  &     0.095    &      0.00            &       0.0   &       0.64     \\
& 1999.893   &     E1 &      0.003    &      1.25$\pm$0.71   &      88.6   &       0.00    \\
& 1999.893   &     E2 &      0.020    &      2.79$\pm$0.17   &      78.3   &       0.83    \\
& 1999.893   &     E3 &      0.006    &      6.95$\pm$0.38   &      82.0    &       0.64    \\
& 1999.893   &     E4 &      0.001    &     13.36$\pm$2.72   &      89.9    &       2.32     \\
& 2003.447   &     E0  &     0.092    &      0.00            &       0.0   &      0.31       \\
& 2003.447   &     E1  &     0.023    &      1.09$\pm$0.082  &      66.1   &      0.64       \\
& 2003.447   &     E2  &     0.015    &      2.80$\pm$0.11   &      80.2   &      1.58       \\
& 2003.447   &     E3  &     0.008    &      5.82$\pm$0.30   &      84.1   &      2.39       \\
& 2003.447   &     E4  &     0.004    &     13.83$\pm$1.21   & 85.0 & 5.35        \\
1133+704 ...... & 1997.055   &     M0  &     0.099    &      0.00            &        0.0  &      0.38     \\
& 1997.055   &     M1  &     0.027    &      0.74$\pm$0.13   &      111.1  &      0.51      \\
& 1997.055   &     M2  &     0.006    &      2.99$\pm$0.50   &       99.0  &      2.23      \\
& 1997.055   &     M3  &     0.005    &      6.23$\pm$0.88   &       76.7  &      4.16       \\
& 1997.055   &     M4  &     0.005    &     15.61$\pm$1.40   &       73.1  &      7.15       \\
& 1998.493   &     M0  &     0.091    &      0.00            &        0.0  &      0.31    \\
& 1998.493   &     M1  &     0.018    &      1.27$\pm$0.12   &      106.9  &      0.70     \\
& 1998.493   &     M2  &     0.008    &      3.34$\pm$0.25   &      100.9  &      1.29      \\
& 1998.493   &     M3  &     0.004    &     14.99$\pm$1.66   &       73.1  &      5.26      \\
& 1998.493   &     M4  &     0.003    &      7.55$\pm$1.66   &       76.0  &      3.42      \\
& 1999.893   &     M0  &     0.080    &      0.00            &        0.0  &      0.35      \\
& 1999.893   &     M1  &     0.003    &      0.92$\pm$0.42   &      119.6  &      0.00      \\
& 1999.893   &     M2  &     0.013    &      2.71$\pm$0.61   &       93.0  &      1.03      \\
& 1999.893   &     M3  &     0.003    &      8.47$\pm$0.26   &       78.8  &      0.75      \\

\hline\hline
\end{longtable}

\clearpage
\begin{table*}
\caption{Proper motions} \label{table2} \centering
\begin{tabular}{c c c c c c c c c c}
\hline\hline
 Object& Component    & $\mu$ ($\rm mas~ yr^{-1}$)     &
$\beta_{\rm app}$ & $\delta$& $\theta$ (deg)&$\Gamma$&$\theta
_{\Gamma= 5}$ (deg)
\\\hline

1133+704  & M1&0.040$\pm$0.015& 0.12$\pm$0.045&     &    &  \\
          &   M2   &  0.18$\pm$0.032          &   0.56$\pm$0.10& &  &    \\
          &     M3 &  0.78$\pm$0.31         &
          2.4$\pm$0.97& 0.8& 41.6  & 4.6&39.9
          \\

1727$+$502& C1& 0.0074$\pm$0.0085& 0.027$\pm$0.031&&  & \\
          &   C2   & 0.032$\pm$0.013           &   0.12$\pm$0.049&&  &       \\
          &     C3 &  0.16$\pm$0.014          &   0.58$\pm$0.051&   &   &   \\
          &    C4  &   0.47$\pm$0.31         &        1.72$\pm$ 1.15&   &  &     \\
          &    C5  &   1.53$\pm$0.26         &
          5.64$\pm$0.96& 1.9 &18.1   &9.6&23.7
          \\
1741$+$196& E1     & 0.043$\pm$0.013      & 0.24$\pm$0.071     & &    & & \\
          &   E2   &   0.0064$\pm$  0.019       &  0.035$\pm$0.11 &  &  &   &  \\
          &     E3 &   0.031$\pm$0.049         &      0.17    $\pm$0.27& & &  &       \\
          &     E4 &    0.23$\pm$0.21        &         1.25  $\pm$1.13 &3.3& 12.3 &2.0 &16.6 \\\hline

\end{tabular}

\end{table*}

\begin{longtable}{llllllllllllllll}

\caption{\label{tab:core}The sample of 170 BL Lac objects selected
from \cite{nieppola2006}. } \\
\toprule IAU name&Source&$z$ & log\,$\rm \nu_{peak}^{'}$
&log\,$P_{\rm 408~ M}$
  &log\,$P_{\rm core}$
&$f_{\rm ext}$&refs.&$\delta$&$\theta_{3}$&$\theta_{5}$ &$\theta_{10}$\\
\midrule \midrule
\endfirsthead  
\midrule IAU name&Source&$z$ & log\,$\rm \nu_{peak}^{'}$
&log\,$P_{\rm 408~ M}$
  &log\,$P_{\rm core}$
&$f_{\rm ext}$&refs.&$\delta$&$\theta_{3}$&$\theta_{5}$ &$\theta_{10}$\\
\midrule
\endhead 

\midrule
\multicolumn{4}{r}{Continued\dots} \\
\endfoot 
\endlastfoot
 0006-063  &  NRAO 5             &      0.347  & 12.10  &  26.94$^{a}$ &     26.65   &    ...        &   1 &    6.0 &    0.0&  9.4  &   8.8      \\
0007+472  &  RX J0007.9+4711    &      0.302* & 15.81  &  25.63$^{b}$ &     25.18   &    ...        &   2 &    2.8 &   21.1& 18.7  &  14.3      \\
0035+598  &  1ES 0033+595       &      0.086  & 18.60  &  24.01$^{c}$ &     24.02   &    4.50       &   3 &    2.3 &   24.7& 21.1  &  15.9      \\
0040+408  &  1ES 0037+405       &      0.271* & 16.62  &  24.93$^{c}$ &     24.42   &     3.07      &   3 &    1.9 &   28.8& 23.9  &  17.7      \\
0050-094  &  PKS 0048-097       &      0.537  & 12.94  &  27.29$^{c}$ &     26.60   &  139.00       &   4 &    4.4 &   11.5& 13.1  &  10.9      \\
0110+418  &  NPM1G +41.0022     &      0.096  & 17.72  &  24.41$^{d}$ &     23.58   &    ...        &   2 &    1.1 &   43.5& 34.2  &  24.5      \\
0112+227  &  S2 0109+22         &      0.473* & 12.83  &  26.36$^{b}$ &     26.60   &     ...       &   2 &    8.5 &    0.0&  4.7  &   6.7      \\
0115+253  &  RXS J0115.7+2519   &      0.350  & 13.15  &  25.31$^{d}$ &     24.92   &    ...        &   2 &    2.6 &   22.5& 19.7  &  14.9      \\
0123+343  &  1ES 0120+340       &      0.272  & 17.96  &  24.86$^{c}$ &     24.75   &    2.59       &   3 &    2.9 &   19.9& 18.0  &  13.9     \\
0124+093  &  MS 0122.1+0903     &      0.339  & 15.37  &  23.58$^{c}$ &     23.60   &    0.08       &   5 &    2.0 &   28.4& 23.6  &  17.5      \\
0136+391  &  B3 0133+388        &      0.271* & 16.31  &  25.44$^{a}$ &     24.95   &     ...       &   2 &    2.4 &   23.7& 20.5  &  15.5      \\
0141-094  &  PKS 0139-09        &      0.733  & 12.77  &  27.17$^{c}$ &     26.99   &   50.00       &   6 &    7.4 &    0.0&  6.7  &   7.5      \\
0148+140  &  1ES 0145+138       &      0.125  & 15.76  &  24.22$^{e}$ &     22.88   &    ...        &   7 &    0.5 &   65.8& 49.9  &  34.9      \\
0153+712  &  8C 0149+710        &      0.022  & 14.65  &  23.99$^{b}$ &     23.50   &    ...        &   2 &    1.3 &   38.2& 30.4  &  22.0      \\
0201+005  &  MS 0158.5+0019     &      0.299  & 17.62  &  24.46$^{c}$ &     24.29   &    0.83       &   5 &    2.3 &   24.9& 21.3  &  16.0      \\
0208+353  &  MS 0205.7+3509     &      0.318  & 14.90  &  23.91$^{c}$ &     24.10   &    0.20       &   5 &    2.7 &   21.2& 18.8  &  14.4      \\
0214+517  &  87GB 02109+5130    &      0.049  & 17.53  &  24.50$^{b}$ &     23.94   &    ...        &   2 &    1.5 &   34.4& 27.8  &  20.3      \\
0222+430  &  3C 66A             &      0.440  & 15.20  &  27.55$^{a}$ &     26.65   &    ...        &   8 &    3.9 &   14.3& 14.6  &  11.8      \\
0232+202  &  1ES 0229+200       &      0.140  & 19.22  &  24.75$^{d}$ &     24.32   &    ...        &   7 &    1.9 &   28.7& 23.8  &  17.7      \\
0238+166  &  AO 0235+164        &      0.940  & 12.83  &  27.24$^{c}$ &     27.35   &   31.40       &   9 &   10.8 &    0.0&  0.0  &   5.3      \\
0301+346  &  MS 0257.9+3429     &      0.245  & 13.10  &  24.49$^{c}$ &     24.14   &    1.38       &   5 &    1.9 &   29.1& 24.1  &  17.8     \\
0314+247  &  RXS J0314.0+2445   &      0.054  & 12.71  &  22.95$^{d}$ &     22.60   &    ...        &   2 &    1.0 &   46.0& 36.0  &  25.7      \\
0326+024  &  2E 0323+0214       &      0.147  & 19.61  &  24.16$^{c}$ &     24.01   &    2.00       &  10 &    2.1 &   27.2& 22.8  &  17.0      \\
0416+010  &  2E 0414+0057       &      0.287  & 20.49  &  25.70$^{d}$ &     24.99   &    ...        &   1 &    2.1 &   26.6& 22.4  &  16.7      \\
0422+198  &  MS 0419.3+1943     &      0.512  & 16.63  &  25.15$^{c}$ &     24.73   &    1.14       &   5 &    2.3 &   24.7& 21.1  &  15.9      \\
0424+006  &  PKS 0422+004       &      0.310  & 15.01  &  26.33$^{b}$ &     26.32   &    ...        &   9 &    6.2 &    0.0&  8.9  &   8.5      \\
0505+042  &  RXS J0505.5+0416   &      0.027  & 16.87  &  23.57$^{d}$ &     23.17   &    ...        &   2 &    1.2 &   39.9& 31.7  &  22.8      \\
0507+676  &  1ES 0502+675       &      0.314  & 18.55  &  24.20$^{c}$ &     24.90   &    0.40       &   3 &    5.6 &    3.8& 10.2  &   9.2      \\
0508+845  &  S5 0454+84         &      0.112  & 12.59  &  23.75$^{c}$ &     25.19   &    5.10$^{f}$ &  4,9&   10.8 &    0.0&  0.0  &   5.3      \\
0509-040  &  4U 0506-03         &      0.304  & 17.77  &  25.57$^{c}$ &     24.82   &   36.00$^{f}$ &  11 &    1.9 &   28.7& 23.8  &  17.7      \\
0613+711  &  MS 0607.9+7108     &      0.267  & 14.41  &  24.03$^{c}$ &     24.39   &    0.40       &   5 &    3.5 &   16.4& 15.8  &  12.5     \\
0625+446  &  87GB 06216+4441    &      0.473* & 13.04  &  26.25$^{d}$ &     26.15   &     ...       &   8 &    5.5 &    5.2& 10.5  &   9.4      \\
0650+250  &  1ES 0647+250       &      0.203  & 17.85  &  24.85$^{c}$ &     24.83   &    4.89       &   3 &    3.2 &   17.9& 16.7  &  13.1      \\
0654+427  &  B3 0651+428        &      0.126  & 14.79  &  25.07$^{a}$ &     24.70   &    ...        &   2 &    2.4 &   24.2& 20.8  &  15.7      \\
0656+426  &  NPM1G +42.0131     &      0.059  & 17.34  &  25.85$^{c}$ &     24.30   &  699.00$^{f}$ &  12 &    0.9 &   49.5& 38.4  &  27.3      \\
0710+591  &  EXO 0706.1+5913    &      0.125  & 20.83  &  25.03$^{b}$ &     24.46   &    ...        &   7 &    1.9 &   29.3& 24.3  &  17.9      \\
0721+713  &  S5 0716+714        &      0.300  & 14.07  &  26.50$^{c}$ &     25.85   &  316.00$^{f}$ &   4 &    3.2 &   18.0& 16.8  &  13.1     \\
0738+177  &  PKS 0735+17        &      0.424  & 12.64  &  25.43$^{c}$ &     27.10   &   12.00$^{f}$ &   4 &   29.4 &    0.0&  0.0  &   0.0      \\
0744+745  &  MS 0737.9+7441     &      0.315  & 13.24  &  24.75$^{c}$ &     24.71   &    1.42       &   5 &    3.1 &   19.1& 17.5  &  13.6      \\
0753+538  &  S4 0749+54         &      0.200  & 12.23  &  25.47$^{e}$ &     26.12   &    ...        &   13&    9.3 &    0.0&  3.1  &   6.2      \\
0757+099  &  PKS 0754+100       &      0.266  & 12.48  &  25.26$^{c}$ &     26.56   &    6.70       &   9 &   17.7 &    0.0&  0.0  &   2.0      \\
0806+595  &  SBS 0802+596       &      0.300  & 16.43  &  25.26$^{e}$ &     24.81   &    ...        &   2 &    2.4 &   24.3& 20.9  &  15.7      \\
0809+523  &  1ES 0806+524       &      0.137  & 16.09  &  24.90$^{e}$ &     24.88   &    ...        &   3 &    3.3 &   17.4& 16.4  &  12.9      \\
0818+423  &  OJ 425             &      0.245  & 12.73  &  26.41$^{a}$ &     26.17   &    ...        &  14 &    5.0 &    8.3& 11.6  &  10.0      \\
0823+223  &  4C 22.21           &      0.951  & 12.94  &  28.53$^{c}$ &     26.96   &   602.60      &   4 &    2.7 &   21.4& 18.9  &  14.5      \\
0825+031  &  PKS 0823+033       &      0.506  & 11.79  &  25.23$^{c}$ &     26.98   &    1.40       &  14 &   29.4 &    0.0&  0.0  &   0.0      \\
0831+044  &  PKS 0829+046       &      0.174  & 12.81  &  25.74$^{b}$ &     25.95   &    ...        &   7 &    6.2 &    0.0&  9.0  &   8.6      \\
0831+087  &  1H 0827+089        &      0.941  & 13.90  &  26.67$^{d}$ &     26.15   &    ...        &  11 &    4.0 &   13.3& 14.0  &  11.4      \\
0832+492  &  OJ 448             &      0.548  & 12.33  &  25.98$^{c}$ &     26.36   &   23.40$^{f}$ &   4 &    8.4 &    0.0&  4.9  &   6.8      \\
0854+441  &  US 1889            &      0.382  & 17.31  &  26.05$^{a}$ &     25.06   &    ...        &   2 &    1.8 &   30.3& 24.9  &  18.4      \\
0854+201  &  OJ 287             &      0.306  & 12.75  &  25.25$^{c}$ &     26.56   &   17.00$^{f}$ &  4,9&   17.9 &    0.0&  0.0  &   2.0      \\
0915+295  &  B2 0912+29         &      0.302* & 15.64  &  26.20$^{b}$ &     25.59   &    ...        &   2 &    3.0 &   19.7& 17.9  &  13.8     \\
0916+526  &  RXS J0916.8+5238   &      0.271* & 17.01  &  25.61$^{d}$ &     24.92   &     ...       &   2 &    2.1 &   27.0& 22.7  &  16.9      \\
0929+502  &  RXS J0929.2+5013   &      0.370  & 13.76  &  26.08$^{e}$ &     26.50   &    ...        &   8 &    9.2 &    0.0&  3.1  &   6.2      \\
0930+498  &  1ES 0927+500       &      0.188  & 20.77  &  24.06$^{e}$ &     24.18   &    ...        &   3&     2.7 &   21.6& 19.1  &  14.6      \\
0930+350  &  B2 0927+35         &      0.302* & 14.35  &  26.48$^{b}$ &     25.95   &     ...       &   8 &    3.7 &   15.3& 15.2  &  12.1      \\
0952+656  &  RGB J0952+656      &      0.302* & 14.90  &  25.46$^{e}$ &     24.78   &     ...       &   2 &    2.0 &   28.0& 23.3  &  17.3      \\
0954+492  &  MS 0950.9+4929     &      0.380  & 16.73  &  24.20$^{c}$ &     24.06   &    0.26       &   5 &    2.1 &   26.6& 22.4  &  16.7      \\
0958+655  &  S4 0954+65         &      0.368  & 13.06  &  25.66$^{c}$ &     25.97   &   28.60$^{f}$ & 4,15&    6.8 &    0.0&  7.9  &   8.0      \\
1012+424  &  RXS J1012.7+4229   &      0.364  & 20.82  &  25.83$^{a}$ &     24.98   &    ...        &   2 &    1.9 &   28.7& 23.9  &  17.7      \\
1015+494  &  GB 1011+496        &      0.200  & 16.41  &  25.80$^{b}$ &     25.22   &    ...        &   16&    2.6 &   22.5& 19.7  &  14.9      \\
1031+508  &  1ES 1028+511       &      0.360  & 18.16  &  25.32$^{e}$ &     25.16   &    ...        &   3 &    3.4 &   17.0& 16.2  &  12.7      \\
1037+571  &  RXS J1037.7+5711   &      0.302* & 14.50  &  25.43$^{d}$ &     25.30   &     ...       &   2 &    3.7 &   15.2& 15.1  &  12.1      \\
1047+546  &  1ES 1044+549       &      0.473* & 12.86  &  24.68$^{d}$ &     24.32   &     ...       &   3 &    2.0 &   27.4& 22.9  &  17.1      \\
1053+494  &  MS 1050.7+4946     &      0.140  & 14.95  &  24.30$^{c}$ &     24.26   &    3.12       &   5 &    2.5 &   23.3& 20.2  &  15.3      \\
1104+382  &  MRK 421            &      0.031  & 18.20  &  24.40$^{a}$ &     24.14   &    ...        &   7&     2.0 &   27.7& 23.2  &  17.2      \\
1109+241  &  1ES 1106+244       &      0.271* & 16.70  &  24.96$^{d}$ &     24.50   &     ...       &   3 &    2.1 &   27.4& 22.9  &  17.1      \\
1120+422  &  EXO 1118.0+4228    &      0.124  & 17.21  &  24.10$^{d}$ &     23.83   &    ...        &   2 &    1.8 &   30.7& 25.2  &  18.6      \\
1136+701  &  MRK 180            &      0.046  & 18.73  &  25.16$^{c}$ &     23.80   &  237.00       &   3 &    0.8 &   52.2& 40.3  &  28.6      \\
1136+676  &  RXS J1136.5+6737   &      0.134  & 17.18  &  24.17$^{e}$ &     24.23   &    ...        &   2 &    2.6 &   22.0& 19.4  &  14.8      \\
1149+246  &  EXO 1449.9+2455    &      0.402  & 19.63  &  25.32$^{d}$ &     24.79   &    ...        &   2 &    2.2 &   25.8& 21.9  &  16.4      \\
1150+242  &  B2 1147+245        &      0.200  & 13.18  &  25.30$^{c}$ &     25.79   &   50.00$^{f}$ & 4,15&    7.1 &    0.0&  7.3  &   7.7     \\
1151+589  &  RXS J1151.4+5859   &      0.302* & 16.10  &  25.98$^{b}$ &     25.33   &     ...       &   2 &    2.6 &   22.6& 19.7  &  15.0      \\
1209+413  &  B3 1206+416        &      0.302* & 13.88  &  25.64$^{a}$ &     25.95   &     ...       &  8  &    6.7 &    0.0&  8.0  &   8.1      \\
1215+075  &  1ES 1212+078       &      0.130  & 15.57  &  24.80$^{d}$ &     24.56   &    ...        &   7 &    2.5 &   23.6& 20.4  &  15.4      \\
1217+301  &  B2 1215+30         &      0.130  & 15.10  &  25.46$^{b}$ &     25.24   &    ...        &   7 &    3.4 &   17.0& 16.2  &  12.7      \\
1220+345  &  GB2 1217+348       &      0.130  & 13.97  &  25.04$^{e}$ &     25.01   &    ...        &   2 &    3.5 &   16.5& 15.9  &  12.5      \\
1221+301  &  PG 1218+304        &      0.182  & 18.80  &  24.86$^{e}$ &     24.65   &    ...        &   7 &    2.6 &   22.4& 19.6  &  14.9      \\
1221+282  &  ON 231             &      0.102  & 14.16  &  25.12$^{c}$ &     25.43   &   41.00       &   3 &    5.3 &    6.2& 10.8  &   9.5      \\
1223+806  &  S5 1221+80         &      0.473* & 13.74  &  26.99$^{e}$ &     26.41   &     ...       &   8 &    4.3 &   11.8& 13.2  &  10.9      \\
1224+246  &  MS 1221.8+2452     &      0.218  & 13.60  &  24.25$^{c}$ &     24.38   &    1.05       &   5 &    3.0 &   19.7& 17.9  &  13.8      \\
1230+253  &  RXS J1230.2+2517   &      0.135  & 14.40  &  25.26$^{b}$ &     25.18   &    ...        &   2 &    3.6 &   15.8& 15.4  &  12.3      \\
1231+642  &  MS 1229.2+6430     &      0.164  & 15.84  &  24.33$^{c}$ &     24.43   &    2.34       &   5 &    3.0 &   19.7& 17.9  &  13.8      \\
1237+629  &  MS 1235.4+6315     &      0.297  & 15.61  &  24.38$^{c}$ &     24.48   &    0.70       &   17&    3.0 &   19.3& 17.6  &  13.6      \\
1241+066  &  1ES 1239+069       &      0.150  & 17.05  &  23.48$^{c}$ &     23.73   &    0.40       &   3 &    2.5 &   23.6& 20.4  &  15.4      \\
1248+583  &  PG 1246+586        &      0.847  & 14.27  &  26.90$^{b}$ &     27.17   &    ...        &   8 &   11.1 &    0.0&  0.0  &   5.1      \\
1253+530  &  S4 1250+53         &      0.302* & 14.39  &  26.45$^{b}$ &     25.89   &     ...       &   8 &    3.5 &   16.2& 15.7  &  12.4      \\
1257+242  &  1ES 1255+244       &      0.141  & 16.83  &  24.00$^{d}$ &     23.51   &    ...        &   3 &    1.3 &   38.0& 30.3  &  22.0      \\
1310+325  &  AUCVn              &      0.996  & 12.61  &  26.54$^{c}$ &     27.74   &    5.40       & 9,15&   27.7 &    0.0&  0.0  &   0.0     \\
1322+081  &  1ES 1320+084N      &      0.049  & 12.99  &  22.74$^{d}$ &     22.81   &    ...        &    3&    1.4 &   35.6& 28.6  &  20.8      \\
1341+399  &  RXS J1341.0+3959   &      0.163  & 19.97  &  25.19$^{a}$ &     24.33   &    ...        &    2&    1.4 &   35.5& 28.5  &  20.8      \\
1402+159  &  MC 1400+162        &      0.244  & 16.29  &  26.54$^{a}$ &     25.41   &    ...        &    1&    1.9 &   29.0& 24.1  &  17.8      \\
1404+040  &  MS 1402.3+0416(went&      0.344  & 15.74  &  25.76$^{c}$ &     24.80   &   11.89       &    3&    1.6 &   32.3& 26.4  &  19.3     \\
1409+596  &  MS 1407.9+5954     &      0.496  & 16.40  &  25.50$^{c}$ &     25.02   &    2.75       &    5&    2.6 &   22.8& 19.8  &  15.1      \\
1415+485  &  RGB J1415+485      &      0.500  & 14.13  &  25.73$^{e}$ &     25.57   &    ...        &    2&    4.1 &   13.1& 14.0  &  11.4      \\
1415+133  &  PKS 1413+135       &      0.247  & 12.39  &  26.66$^{a}$ &     26.01   &    ...        &    1&    3.5 &   16.5& 15.9  &  12.6      \\
1417+257  &  2E 1415+2557       &      0.237  & 19.00  &  25.29$^{d}$ &     24.74   &    ...        &    2&    2.1 &   26.5& 22.4  &  16.7      \\
1419+543  &  OQ 530             &      0.152  & 13.17  &  24.68$^{c}$ &     25.81   &   22.00$^{f}$ &  4,7&   11.4 &    0.0&  0.0  &   5.0      \\
1427+238  &  PKS 1424+240       &      0.160  & 15.34  &  25.69$^{a}$ &     25.18   &    ...        &    2&    2.7 &   21.9& 19.3  &  14.7      \\
1427+541  &  RGB J1427+541      &      0.106  & 14.79  &  24.38$^{e}$ &     23.80   &    ...        &    2&    1.4 &   36.4& 29.2  &  21.2      \\
1428+426  &  H 1426+428         &      0.129  & 18.41  &  24.43$^{e}$ &     23.93   &    ...        &    7&    1.6 &   33.5& 27.2  &  19.9      \\
1439+395  &  PG 1437+398        &      0.344  & 16.56  &  25.98$^{a}$ &     25.05   &    ...        &    2&    1.9 &   29.4& 24.3  &  18.0      \\
1442+120  &  1ES 1440+122       &      0.162  & 16.20  &  24.80$^{d}$ &     24.41   &    ...        &    3&    2.1 &   27.3& 22.9  &  17.0      \\
1444+636  &  MS 1443.5+6349     &      0.298  & 16.94  &  24.31$^{c}$ &     23.93   &    0.59       &    5&    1.7 &   31.6& 25.9  &  19.0      \\
1448+361  &  RXS J1448.0+3608   &      0.271* & 16.43  &  25.03$^{d}$ &     24.72   &     ...       &    2&    2.5 &   23.1& 20.1  &  15.2      \\
1458+373  &  B3 1456+375        &      0.333  & 12.86  &  25.91$^{a}$ &     25.93   &    ...        &    8&    5.4 &    5.7& 10.7  &   9.5      \\
1501+226  &  MS 1458.8+2249     &      0.235  & 14.69  &  24.73$^{c}$ &     25.06   &    2.66       &    5&    4.6 &   10.4& 12.5  &  10.5     \\
1509+559  &  SBS 1508+561       &      0.302* & 14.83  &  24.83$^{e}$ &     24.80   &     ...       &    2&    3.2 &   18.3& 17.0  &  13.2      \\
1516+293  &  RXS J1516.7+2918   &      0.130  & 18.56  &  25.01$^{a}$ &     24.13   &    ...        &    2&    1.3 &   38.1& 30.4  &  22.0      \\
1517+654  &  1H 1515+660        &      0.702  & 17.82  &  25.72$^{e}$ &     25.39   &    ...        &    2&    3.3 &   17.4& 16.4  &  12.9      \\
1532+302  &  RXS J1532.0+3016   &      0.064  & 16.86  &  23.87$^{d}$ &     23.64   &    ...        &    2&    1.7 &   32.0& 26.1  &  19.2      \\
1533+342  &  RXS J1533.4+3416   &      0.810  & 17.89  &  25.78$^{e}$ &     25.75   &    ...        &    2&    4.9 &    8.9& 11.9  &  10.1      \\
1534+372  &  RGB J1534+372      &      0.143  & 13.97  &  24.02$^{e}$ &     23.98   &    ...        &    2&    2.2 &   25.8& 21.8  &  16.3      \\
1535+533  &  1ES 1533+535       &      0.890  & 19.55  &  25.96$^{d}$ &     25.31   &    ...        &    3&    2.5 &   22.8& 19.9  &  15.1      \\
1536+016  &  MS 1534.2+0148     &      0.312  & 18.67  &  25.53$^{c}$ &     24.79   &    8.78       &    5&    1.9 &   29.1& 24.1  &  17.8      \\
1540+819  &  1ES 1544+820       &      0.271* & 17.53  &  25.42$^{e}$ &     24.89   &     ...       &    3&    2.3 &   24.9& 21.3  &  16.0      \\
1540+147  &  4C 14.6            &      0.605  & 14.25  &  27.58$^{c}$ &     27.10   &  202.00       &    9&    6.3 &    0.0&  8.7  &   8.4      \\
1542+614  &  RXS J1542.9+6129   &      0.302* & 14.19  &  25.27$^{e}$ &     25.36   &    ...        &    2&    4.4 &   11.3& 13.0  &  10.8      \\
1554+201  &  MS 1552.1+2020     &      0.222  & 16.89  &  25.09$^{c}$ &     24.58   &    6.86       &    5&    2.1 &   27.3& 22.9  &  17.0     \\
1555+111  &  PG 1553+11         &      0.360  & 15.92  &  26.29$^{b}$ &     26.11   &    ...        &    2&    5.1 &    7.7& 11.4  &   9.9      \\
1602+308  &  RXS J1602.2+3050   &      0.302* & 16.27  &  25.36$^{d}$ &     24.65   &     ...       &    2&    1.8 &   29.6& 24.5  &  18.1      \\
1626+352  &  RXS J1626.4+3513   &      0.497  & 15.05  &  25.39$^{d}$ &     24.95   &    ...        &    2&    2.5 &   23.0& 20.0  &  15.1      \\
1644+457  &  RXS J1644.2+4546   &      0.225  & 17.28  &  25.67$^{a}$ &     24.89   &    ...        &    2&    1.9 &   28.4& 23.7  &  17.5      \\
1652+403  &  RGB J1652+403      &      0.240  & 14.80  &  24.60$^{d}$ &     24.19   &    ...        &    2&    1.8 &   29.6& 24.5  &  18.1      \\
1653+397  &  MRK 501            &      0.034  & 16.17  &  23.79$^{c}$ &     24.51   &   67.00$^{f}$ &  4,7&    4.8 &    9.3& 12.0  &  10.2      \\
1704+716  &  RXS J1704.8+7138   &      0.350  & 15.32  &  25.06$^{e}$ &     24.72   &    ...        &   2 &    2.5 &   23.6& 20.4  &  15.4      \\
1719+177  &  PKS 1717+177       &      0.137  & 12.43  &  25.19$^{a}$ &     25.42   &    ...        &   2 &    5.0 &    7.9& 11.5  &   9.9      \\
1724+400  &  B2 1722+40         &      1.049  & 12.64  &  27.81$^{a}$ &     26.99   &    ...        &   8 &    4.7 &    9.7& 12.2  &  10.3      \\
1725+118  &  H 1722+119         &      0.018  & 15.75  &  23.06$^{d}$ &     22.80   &    ...        &   2 &    1.1 &   41.6& 32.9  &  23.6      \\
1728+502  &  IZw187             &      0.055  & 17.16  &  24.20$^{b}$ &     23.96   &    ...        &  10 &    1.9 &   28.9& 24.0  &  17.8      \\
1739+476  &  OT 465             &      0.473* & 13.34  &  27.13$^{b}$ &     26.69   &     ...       &   8 &    5.4 &    5.5& 10.6  &   9.4      \\
1742+597  &  RGBJ 1742+597      &      0.400  & 13.75  &  25.72$^{e}$ &     25.49   &    ...        &   2 &    3.7 &   15.0& 15.0  &  12.0      \\
1743+195  &  NPM1G +19.0510     &      0.084  & 17.44  &  24.40$^{c}$ &     24.53   &   11.60       &   3 &    3.2 &   18.2& 17.0  &  13.2      \\
1745+398  &  B3 1743+398B       &      0.267  & 17.59  &  26.55$^{a}$ &     25.31   &    ...        &   2 &    1.7 &   31.6& 25.9  &  19.0      \\
1747+469  &  B3 1746+470        &      1.484  & 13.30  &  27.26$^{a}$ &     27.41   &    ...        &   8 &   11.3 &    0.0&  0.0  &   5.0      \\
1748+700  &  S4 1749+70         &      0.770  & 13.80  &  26.63$^{c}$ &     26.91   &   45.70$^{f}$ & 4,14&    9.9 &    0.0&  0.0  &   5.8      \\
1749+433  &  B3 1747+433        &      0.473* & 13.08  &  26.62$^{a}$ &     26.21   &    ...        &   8 &    4.5 &   11.1& 12.9  &  10.7      \\
1750+470  &  RXS J1750.0+4700   &      0.160  & 18.31  &  25.28$^{a}$ &     23.78   &    ...        &   2 &    0.7 &   55.5& 42.7  &  30.2      \\
1751+096  &  PKS 1749+096       &      0.322  & 11.33  &  24.92$^{c}$ &     26.99   &    2.00       &  14 &   37.1 &    0.0&  0.0  &   0.0     \\
1756+553  &  RXS J1756.2+5522   &      0.271* & 19.74  &  24.72$^{e}$ &     24.26   &     ...       &    2&    1.8 &   29.7& 24.6  &  18.1      \\
1757+705  &  MS 1757.7+7034     &      0.407  & 13.37  &  24.65$^{c}$ &     24.64   &    0.62       &    5&    3.0 &   19.3& 17.6  &  13.6      \\
1800+784  &  S5 1803+784        &      0.680  & 13.35  &  27.57$^{e}$ &     27.36   &    ...        &4,15 &    8.5 &    0.0&  4.7  &   6.7      \\
1806+698  &  3C 371             &      0.051  & 14.42  &  25.87$^{c}$ &     24.95   &   990.00      & 4,7 &    1.8 &   30.4& 25.0  &  18.4      \\
1808+468  &  RGB J1808+468      &      0.450  & 14.34  &  25.85$^{e}$ &     25.33   &    ...        &   2 &    2.8 &   20.8& 18.6  &  14.3      \\
1811+442  &  RGB J1811+442      &      0.350  & 15.30  &  25.92$^{e}$ &     24.26   &    ...        &   2 &    0.8 &   52.3& 40.4  &  28.7      \\
1813+317  &  B2 1811+31         &      0.117  & 15.34  &  25.00$^{b}$ &     24.37   &    ...        &   2 &    1.7 &   31.1& 25.5  &  18.8      \\
1824+568  &  4C 56.27           &      0.664  & 12.56  &  28.02$^{c}$ &     26.99   &  452.00       &   9 &    4.1 &   13.2& 14.0  &  11.4      \\
1829+540  &  RXS J1829.4+5402   &      0.302* & 15.06  &  25.73$^{e}$ &     24.61   &     ...       &   2 &    1.3 &   37.1& 29.7  &  21.5      \\
1838+480  &  RXS J1838.7+4802   &      0.300  & 13.24  &  25.05$^{d}$ &     24.71   &    ...        &   2 &    2.5 &   23.6& 20.4  &  15.4      \\
1841+591  &  RGB J1841+591      &      0.530  & 14.94  &  25.68$^{e}$ &     24.64   &    ...        &   2 &    1.4 &   35.4& 28.5  &  20.7      \\
1853+672  &  1ES 1853+671       &      0.212  & 16.25  &  23.81$^{c}$ &     24.12   &    0.40       &   3 &    3.0 &   19.5& 17.7  &  13.7      \\
1927+612  &  S4 1926+61         &      0.473* & 12.72  &  26.62$^{b}$ &     26.67   &     ...       &  8  &    7.7 &    0.0&  6.3  &   7.3      \\
1959+651  &  1ES 1959+650       &      0.047  & 17.33  &  23.01$^{c}$ &     24.10   &    1.60       &  3  &    5.2 &    6.9& 11.1  &   9.7      \\
2005+778  &  S5 2007+77         &      0.342  & 12.39  &  26.14$^{c}$ &     26.38   &   28.90       &  9  &    7.7 &    0.0&  6.2  &   7.3      \\
2009+724  &  S5 2010+72         &      0.473* & 12.94  &  27.02$^{e}$ &     26.90   &     ...       &  8  &    7.5 &    0.0&  6.7  &   7.5      \\
2022+761  &  S5 2023+76         &      0.473* & 13.54  &  26.65$^{e}$ &     26.39   &     ...       &  13 &    5.4 &    5.7& 10.7  &   9.5      \\
2039+523  &  1ES 2037+521       &      0.053  & 15.60  &  22.27$^{c}$ &     23.31   &    0.23       & 3   &    3.6 &   16.0& 15.6  &  12.4      \\
2134-018  &  PKS 2131-021       &      1.285  & 12.04  &  27.93$^{c}$ &     27.87   &   70.80       &  14 &   11.9 &    0.0&  0.0  &   4.7      \\
2145+073  &  MS 2143.4+0704     &      0.237  & 13.61  &  25.13$^{c}$ &     24.79   &    6.60       &   5 &    2.5 &   22.9& 20.0  &  15.1      \\
2152+175  &  PKS 2149+17        &      0.473* & 13.00  &  26.03$^{c}$ &     26.57   &    36.70$^{f}$&   4 &   10.4 &    0.0&  0.0  &   5.5      \\
2202+422  &  BL LAC             &      0.070  & 13.15  &  24.21$^{c}$ &     25.73   &   40.00$^{f}$ & 4,7 &   14.5 &    0.0&  0.0  &   3.5      \\
2250+384  &  B3 2247+381        &      0.119  & 15.35  &  24.64$^{a}$ &     24.29   &    ...        &  2  &    2.0 &   27.6& 23.1  &  17.2      \\
2257+077  &  PKS 2254+074       &      0.190  & 13.31  &  24.78$^{c}$ &     25.66   &   17.00$^{f}$ &9,10 &    8.9 &    0.0&  4.0  &   6.4      \\
2319+161  &  Q J2319+161        &      0.302* & 15.29  &  25.13$^{d}$ &     24.58   &     ...       &    2&    2.0 &   27.9& 23.3  &  17.3      \\
2322+346  &  TEX 2320+343       &      0.098  & 16.68  &  24.59$^{b}$ &     23.82   &   ...         &    2&    1.2 &   39.4& 31.3  &  22.6      \\
2323+421  &  1ES 2321+419       &      0.059  & 13.05  &  23.09$^{d}$ &     23.18   &    ...        &   11&    1.7 &   31.5& 25.8  &  18.9      \\
2329+177  &  1ES 2326+174       &      0.213  & 17.84  &  24.64$^{d}$ &     24.30   &    ...        &   3 &    2.0 &   27.4& 23.0  &  17.1      \\
2339+055  &  MS 2336.5+0517     &      0.740  & 14.89  &  25.74$^{d}$ &     24.88   &    ...        &   5 &    1.8 &   30.0& 24.7  &  18.2     \\
2347+517  &  1ES 2344+514       &      0.044  & 15.86  &  23.30$^{c}$ &     23.97   &    3.58       &   3 &    3.6 &   15.6& 15.3  &  12.2      \\
2350+196  &  MS 2347.4+1924     &      0.515  & 15.72  &  24.80$^{d}$ &     24.29   &    ...        &   5 &    1.8 &   30.0& 24.8  &  18.3      \\

\bottomrule

\end{longtable}
\vspace{-4mm} Notes: Col. 1: the source IAU name (J2000). Col. 2:
the source alias name. Col. 3: the redshift. `*' indicate that the
redshift is unknown, and taken as the average redshift of
LBL/IBL/HBL subclass. Col. 4: the intrinsic synchrotron peak
frequency after excluding the Doppler factor. Col. 5: the total
radio power at 408 MHz (notes: `a' - from 408 MHz; `b' - from 360
MHz; `c' - from extended flux; `d' - from total 1.4 GHz; `e' - from
330 MHz). Col. 6: the 5 GHz core luminosity. Col. 7: the 5 GHz
extended flux density in mJy, `f' indicates the extend flux at 1.4
GHz. Col. 8: the references for core and extended flux: ( 1:
\cite{liuandzhang}
 2:     \cite{laure99}
 3:    \cite{per96}
 4:     \cite{rec01}
 5:    \cite{cavallotti04}
 6:     \cite{cas99}
 7:     \cite{gir04b}
 8:     \cite{Tay96}
 9:    \cite{mur93}
 10:     \cite{lau93}
 11:     \cite{koll}
 12:    \cite{rec03}
 13:    \cite{taylor94}
 14:    \cite{re03}
 15:    \cite{cas02}
 16:     \cite{AWB98}
17:     \cite{rec00} ). Col. 9: the Doppler factor. Cols. 10 - 12:
the viewing angle of $\Gamma$= 3, 5 and 10, respectively.

\clearpage


\end{document}